\documentclass{article}

\usepackage{xcolor}
\usepackage{arxiv}

\usepackage[utf8]{inputenc} 
\usepackage[T1]{fontenc}    
\usepackage{hyperref}       
\usepackage{url}            
\usepackage{booktabs}       
\usepackage{amsfonts}       
\usepackage{nicefrac}       
\usepackage{microtype}      
\usepackage{lipsum}		
\usepackage{graphicx}
\usepackage[square,sort,comma,numbers]{natbib}
\usepackage{doi}
\usepackage{amsmath,amssymb,amsfonts}
\usepackage[linesnumbered,ruled,vlined]{algorithm2e}
\SetKwInput{KwInput}{Input}                
\SetKwInput{KwOutput}{Output}              
\usepackage[noend]{algpseudocode}
\usepackage{dsfont}
\usepackage{fixltx2e}
\usepackage{multirow} 
\newcommand{\beginsupplement}{%
        \setcounter{table}{0}
        \renewcommand{\thetable}{S\arabic{table}}%
        \setcounter{figure}{0}
        \renewcommand{\thefigure}{S\arabic{figure}}%
        \setcounter{algocf}{0}
        \renewcommand{\thealgocf}{S\arabic{algocf}}
        \setcounter{section}{0}
        \renewcommand{\thesection}{S\arabic{section}}
     }
\title{Unconditional Latent Diffusion Models Memorize Patient Imaging Data: Implications for Openly Sharing Synthetic Data}

\author{Salman Ul Hassan Dar$^{1,2,3}$ \And
Marvin Seyfarth$^{1}$ \And
Isabelle Ayx $^{4}$\And
Theano Papavassiliu$^{2,3,5}$\And
Stefan O. Schoenberg $^{2,4}$ \And
Robert Malte Siepmann$^{6}$  \And
Fabian Christopher Laqua$^{7}$  \And
Jannik Kahmann$^{4}$ \And
Norbert Frey$^{1,3}$ \And
Bettina Baeßler$^{7}$ \And
Sebastian Foersch$^{8}$ \And
Daniel Truhn$^{6}$  \And
Jakob Nikolas Kather$^{9,10,11}$   \And
Sandy Engelhardt$^{1,2,3}$ \And \\
    $^{1}$Department of Internal Medicine III, Heidelberg University Hospital, Germany \\
    $^{2}$AI Health Innovation Cluster (AIH), Germany \\
    $^{3}$German Centre for Cardiovascular Research (DZHK), Partner site Heidelberg/Mannheim, Germany \\
    $^{4}$Department of Radiology and Nuclear Medicine, University Medical Center Mannheim, Germany \\
    $^{5}$Department of Cardiology, Angiology, Hemostasis, and Medical Intensive Care, \\University Medical Centre Mannheim, Medical Faculty Mannheim, University of Heidelberg, Germany.\\    
    $^{6}$Diagnostic and Interventional Radiology, University Hospital Aachen, Aachen, Germany\\
    $^{7}$Department of Diagnostic and Interventional Radiology, University Hospital Würzburg, Germany.\\
    $^{8}$Institute of Pathology, University Medical Center Mainz, Mainz, Germany\\
    $^{9}$Else Kroener Fresenius Center for Digital Health, Medical Faculty Carl Gustav Carus, \\TUD Dresden University of Technology, Dresden, Germany\\ 
    $^{10}$Department of Medicine I, University Hospital Dresden, Dresden, Germany\\
    $^{11}$Medical Oncology, National Center for Tumor Diseases (NCT), University Hospital Heidelberg, \\Heidelberg, Germany
}

\renewcommand{\shorttitle}{Memorization in Unconditional Diffusion Models}

\begin{document}
\maketitle

\begin{abstract}
AI models present a wide range of applications in the field of medicine. However, achieving optimal performance requires access to extensive healthcare data, which is often not readily available. Furthermore, the imperative to preserve patient privacy restricts patient data sharing with third parties and even within institutes. Recently, generative AI models have been gaining traction for facilitating open-data sharing by proposing synthetic data as surrogates of real patient data. Despite the promise, \textcolor{black}{some of} these models are susceptible to patient data memorization, where models generate patient data copies instead of novel synthetic samples. This undermines the whole purpose of preserving patient data privacy and may even result in patient re-identification. Considering the importance of the problem, surprisingly it has received relatively little attention in the medical imaging community. To this end, we assess memorization in unconditional latent diffusion models, which are the building blocks of some of the most advanced generative AI models. We train 2D and 3D latent diffusion models on CT, MR, and X-ray datasets for synthetic data generation. Afterwards, we detect the amount of training data memorized utilizing our \textcolor{black}{novel self-supervised copy detection} approach and further investigate various factors that can influence memorization by training models in different settings. Our findings show a surprisingly high degree of patient data memorization across all datasets, with approximately \textcolor{black}{37.2}\% of patient data detected as memorized and \textcolor{black}{68.7}\% of synthetic samples identified as patient data copies on average in our experiments. \textcolor{black}{Comparison with non-diffusion generative models, such as autoencoders and generative adversarial networks, indicates that while latent diffusion models are more susceptible to memorization, overall they outperform non-diffusion models in synthesis quality.} Further analyses reveal that using augmentation strategies during training, \textcolor{black}{small architecture size, and increasing dataset} can reduce memorization while over-training the models can enhance it. Collectively, our results emphasize the importance of carefully training generative models on private medical imaging datasets, and examining the synthetic data to ensure patient privacy before sharing it for medical research and applications.
\end{abstract}


\section{Introduction} 
\begin{figure*}[t]
\centering
\includegraphics[width=1\linewidth]{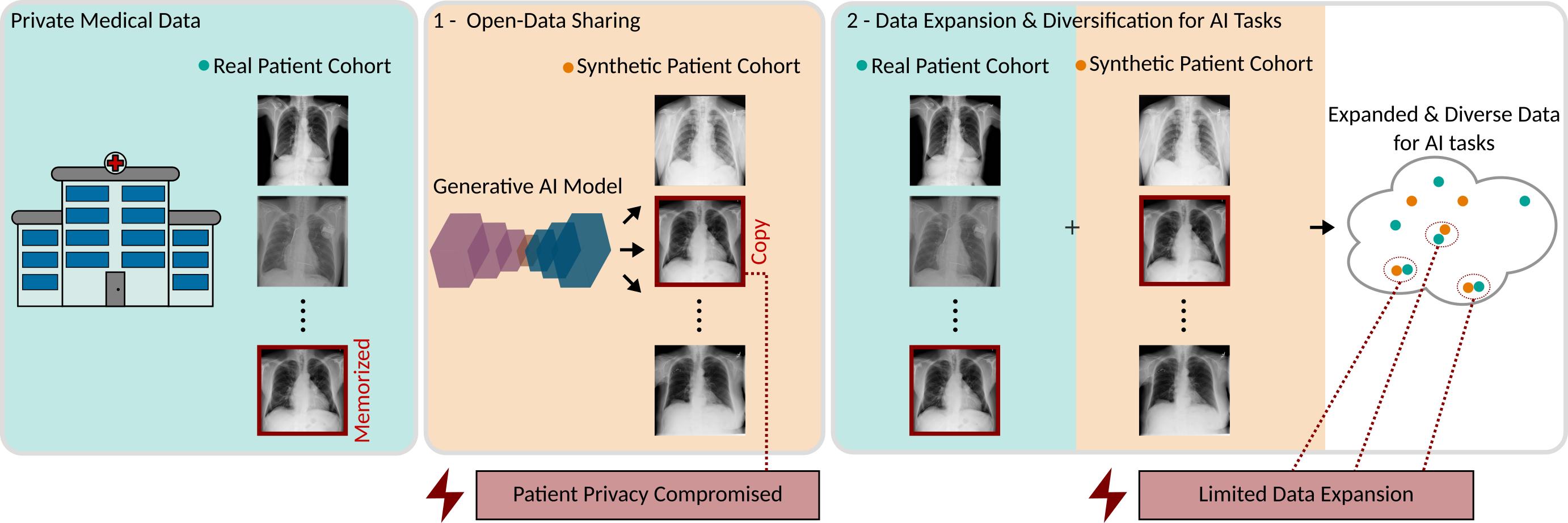}
\caption{Generative models are first trained on private medical data. These models can be used to synthesize novel samples, which can have multiple applications. 1) Open-data sharing: Synthesized samples can be shared publicly for advancing medical imaging research while preserving patient privacy. However, synthesized samples can be patient data replicas, thereby compromising patient privacy. 2) Data Expansion and Diversification - Synthetic samples can be utilized to expand and diversify the training data. Nevertheless, if most of the synthetic samples are patient data replicas, the expansion and diversification is likely to be limited.} 
\label{fig:main_fig}
\end{figure*}

Recent developments in artificial intelligence (AI) hold the potential to transform the current healthcare system. AI models are generally data-hungry, necessitating the healthcare data to scale with the contemporary AI models. While this could be alleviated via patient data sharing among multiple imaging sites and research centers, concerns regarding patient privacy render it infeasible. Modern developments in generative deep diffusion probabilistic models have led to a significant leap in performance level in various medical imaging applications \cite{datadiversification,diffusionrecon,diffusionanomaly,diffusion4medreview}, a notable application being open data sharing \cite{Pinaya2022,opendata1,opendata2,opendata3,opendata4}.
In open-data sharing, generative models are first trained to learn data distribution from private medical imaging datasets. Afterwards, these generative models are used to generate synthetic samples, and since these synthesized samples do not belong to any specific patient, they can be shared publicly without compromising patient privacy (Fig. \ref{fig:main_fig}).  
As a matter of fact, very recently, several studies have trained generative models on private/limited-access/restricted datasets and made synthetic data \cite{Pinaya2022} or trained generative models \cite{hamamci2023generatect,roentgen} publicly available. \\
Despite the potential of generative models for open data sharing, an underlying assumption is that the generated samples are novel and not mere patient data replicas. 
This is crucial, as the primary motivation for using synthetic data as surrogates of real patient data is to preserve patient privacy and synthesizing patient data copies circumvents the goal.
A synthesized copy can even be traced back to the original patient, leading to patient re-identification \cite{packhauser2022reidentification}.
Given the sensitive nature of patient medical data, surprisingly there has been little focus on the threat of such models to memorize training data and detect those memorized training samples efficiently.\\
Detecting memorized samples in the training data can be challenging. Identifying whether a sample is memorized requires comparing it with all synthesized samples, which is sub-optimal both in terms of computational complexity and detection performance. For instance, if a synthesized sample is a slightly rotated copy of a training sample, their pixel-wise differences could still be significant and the patient data copy might go undetected. 
For this purpose, copy detection can be performed via self-supervised models trained based on contrastive learning. In such models, copy detection is performed in a low dimensional embedding space, which makes the whole process computationally efficient and further enables the detection of copies among the synthesized samples that are variants such as rotated versions of the training samples. Such models have been demonstrated for patient re-identification and copy detection in 2D X-ray images \cite{packhauser2022reidentification,dar2024epoch}. However, such applications in 3D medical images have yet to be demonstrated.  \\
While training generative models, the emphasis is typically on improving validation errors or metrics that quantify image quality \cite{FID} or diversity \cite{MS-SSIM}, without taking into consideration the memorization capacities of such models. 
Despite their widespread usage, the commonly used metrics have inherent limitations \cite{Borji2022metrics,jayasumana2024rethinking,chong2020effectively}, and give no direct information regarding patient data memorization. Likewise, the validation loss itself only provides auxiliary information regarding model training and can even show a negative correlation with data memorization \cite{MemorizationVsOverfitting}.
This also makes it challenging to determine an appropriate number of training steps.
In fact, over-training is one of the several factors that could influence memorization, and other factors like training data size and data augmentation can also have an impact on memorization \cite{dar2023investigating,Somepalli2023}.
Therefore, exploring memorization-informed model training and metrics is crucial \cite{Borji2022metrics}. \\
Here, we thoroughly investigate memorization in unconditional latent diffusion models (LDMs) for medical imaging. LDMs learn data generation in the low-dimensional latent space of an autoencoder, which makes them computationally efficient while preserving high image quality \cite{LDM}. \textcolor{black}{This capability is essential in medical images, as medical images are often high-dimensional, and training some generative models at large scales might be challenging. Furthermore, belonging to the family of generative diffusion models, LDMs offer more stability during training compared to other leading generative models such as generative adversarial networks (GANs), which can be difficult to train \cite{diffbeatgans,diffbeatganmed}. In addition, LDMs also} form the foundation of advanced multi-modal and conditional generative AI tools such as stable diffusion, which typically adapt pre-trained unconditional LDMs or perform hybrid training with unconditional LDMs \cite{LDM,controlnet}. 
We train unconditional LDMs on medical images to learn data distributions and perform patient data copy detection among the synthesized samples using self-supervised models. \textcolor{black}{Furthermore, to explore the phenomenon of memorization in other generative models compared to LDMs, we also trained GANs and autoencoder transformers, evaluating their memorization behavior relative to LDMs.}\\
As a means to understand memorization in unconditional diffusion models, we pose the following questions regarding memorization in LDMs for medical image synthesis:
\paragraph{Prevalence:}\vspace{-1em}Is memorization equally prevalent in 2D and 3D LDMs, as well as in medical images with varying properties such as organs, dimension, resolution, field-of-view, contrast, and modality? (section \ref{results:memorization})
\textcolor{black}{\paragraph{Comparison to other Generative Models:}\vspace{-1em} Is memorization also prevalent in other generative models such as autoencoder-transformers and GANs? (section \ref{results:mem_gans_ae})}
\paragraph{\textcolor{black}{Accurate} Detection:}\vspace{-1em}How \textcolor{black}{accurately} can patient data memorization be detected? (section \ref{results:copy_detection_perf})
\textcolor{black}{\paragraph{Robust Detection:}\vspace{-1em}How robust is the proposed copy detection approach? (section \ref{results:copy_det_reobust})}
\textcolor{black}{\paragraph{Quality of the Detected Copies:}\vspace{-1em} How similar are the detected copies to the corresponding training samples? (section \ref{results:rad_eval_copies})}
\paragraph{Impact of Training Data Size:}\vspace{-1em}How is memorization affected by the training data size? (section \ref{results:train_data_size})
\paragraph{Memorization as a Metric:}\vspace{-1em}Can memorization be used as a metric to assess generative models during training? (section \ref{results:iterations})
\paragraph{Comparison with Traditional Metrics:}\vspace{-1em}Can a link between memorization and traditional metrics for assessing generative models be established? (section \ref{results:iterations})
\paragraph{Mitigation via Data Augmentation:}\vspace{-1em}Can data memorization be alleviated by data augmentation during model training? (section \ref{results:augmentation})
\textcolor{black}{\paragraph{Impact of Model Size:}\vspace{-1em}Can data memorization be mitigated by reducing model size? (section \ref{results:architecture})}
\section{Results}
\subsection{Experimental Settings}
\paragraph{Datasets}
We performed a thorough assessment of memorization in unconditional latent diffusion models (LDMs) by conducting analyses on three medical imaging datasets covering a range of modalities, organs, image resolution, fields of view, and spatial dimensions. We conducted experiments on 3D volumes from a publicly available knee MRI dataset (MRNet) \cite{MRNet}, 3D sub-volumes surrounding plaques from an in-house photon counting coronary computed tomography angiography dataset (PCCTA), \textcolor{black}{3D volumes from a publicly available brain MRI dataset (fastMRI) \cite{fastmri1,fastmri2}}, and 2D images from a publicly available X-ray dataset \cite{NIHXray}. In the MRNet dataset, 904 volumes were used for training and 226 for validation. In the PCCTA dataset, 242 volumes were used for training and 58 for validation. \textcolor{black}{In the fastMRI dataset, 1k images were used for training and 1k for validation}. In the X-ray dataset, 10k images were reserved for training and 10k for validation.
\paragraph{Generative Models}
For 3D datasets, Medical Diffusion (MedDiff) \cite{Khader2023} and Medical Open Network
for Artificial Intelligence (MONAI) \cite{pinaya2023monai} based LDMs were adapted. For the X-ray dataset, a Medical Open Network
for Artificial Intelligence (MONAI-2D) based LDM was adapted. LDMs were trained on all datasets separately. Afterward, each model was used to synthesize novel samples. These synthesized samples were then categorized as novel or copies using self-supervised models (see section \ref{methods:mem_assesment} for details). We opted for MedDiff and MONAI because they are based on some of the most widely used repositories for LDMs-based medical image synthesis.
\paragraph{Memorization}
We considered two aspects of memorization. \\
1 - First, we assess the \textbf{number of training samples memorized}. This constitutes the number of training samples that are synthesized as patient data replicas among the synthetic samples ($N_{mem}$).\\
2 - Second, we look at the \textbf{number of synthetic samples that are patient data replicas} ($N_{copies}$). \textcolor{black}{Theoratically, }$N_{copies}$ should be always greater than or equal to $N_{mem}$ because a patient data replica can be repeated multiple times in the synthesized data. 
\subsection{\textcolor{black}{Prevalance of }Memorization}\label{results:memorization}
\begin{figure}[h]
\centering
\includegraphics[width=0.9\linewidth]{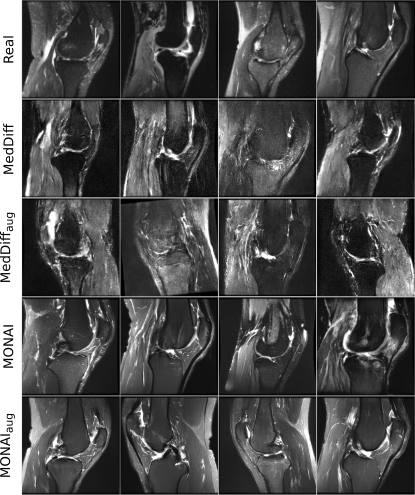}
\caption{Central slices of randomly selected real samples are shown along with central slices of randomly selected MedDiff-synthesized, MedDiff\textsubscript{aug}-synthesized, MONAI-synthesized, and MONAI\textsubscript{aug}-synthesized samples. MedDiff-synthesized samples are noisy and are unable to produce finer structural details. MedDiff\textsubscript{aug}-synthesized samples are noisier and not realistic. MONAI-synthesized and MONAI\textsubscript{aug}-synthesized samples are more realistic but are blurry.} 
\label{suppfig:mrnet_slices}
\end{figure}

\begin{figure}[h]
\centering
\includegraphics[width=1\linewidth]{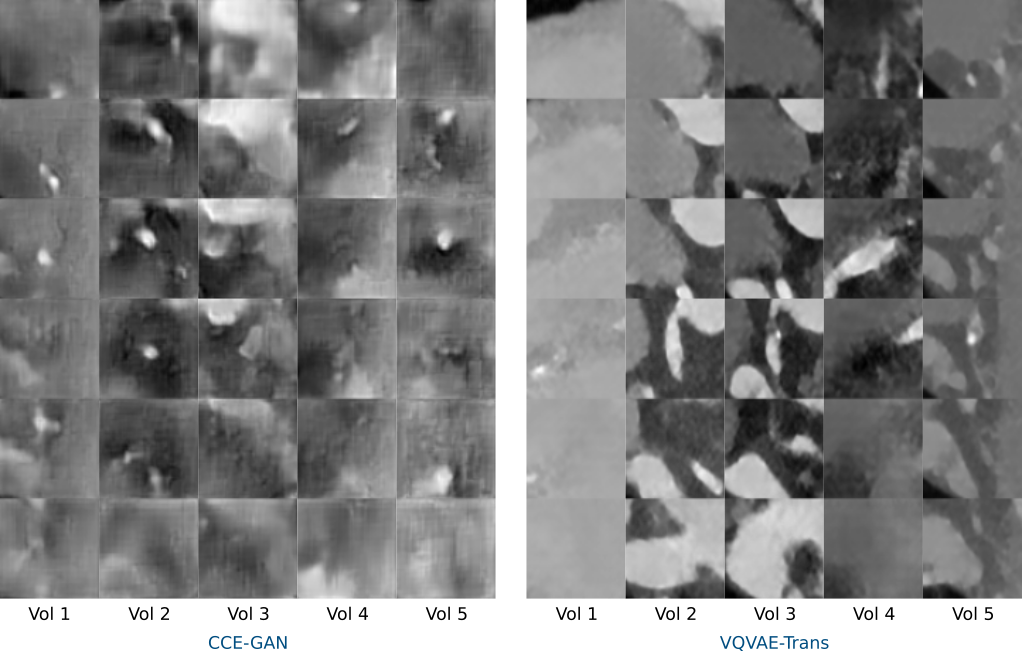}
\caption{Representative volumes synthesized by CCE-GAN and VQVAE-Trans in the PCCTA dataset. CCE-GAN is unable to synthesize realistic images.} 
\label{suppfig:nondiff_pccta}
\end{figure}

\begin{figure}[h]
\centering
\includegraphics[width=1\linewidth]{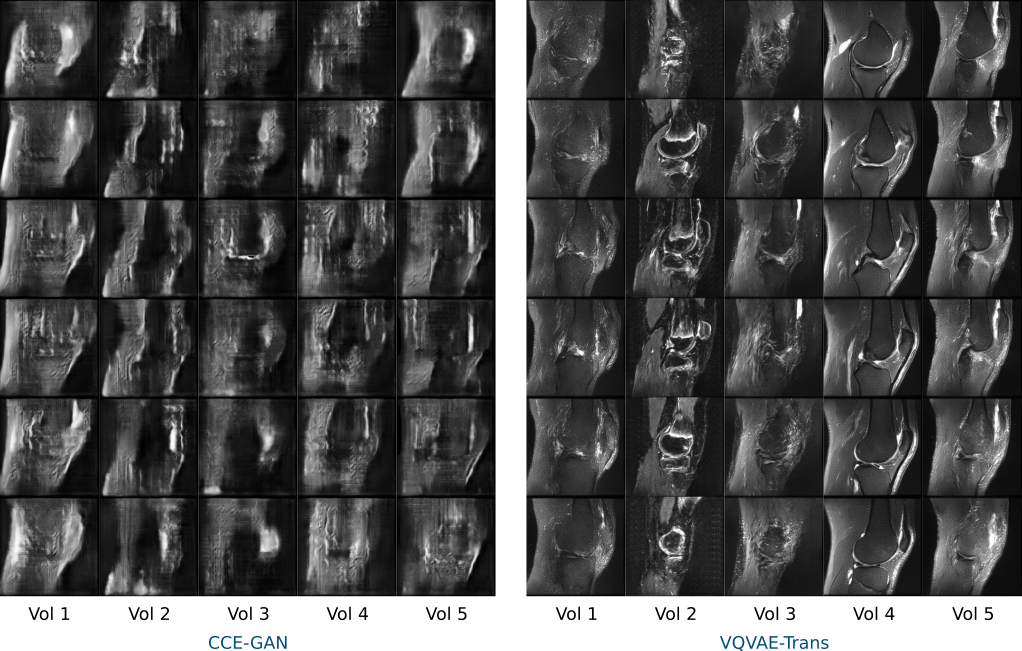}
\caption{Representative volumes synthesized by CCE-GAN and VQVAE-Trans in the MRNet dataset. CCE-GAN is unable to synthesize realistic images.} 
\label{suppfig:nondiff_mrnet}
\end{figure}

\begin{figure}[h]
\centering
\includegraphics[width=1\linewidth]{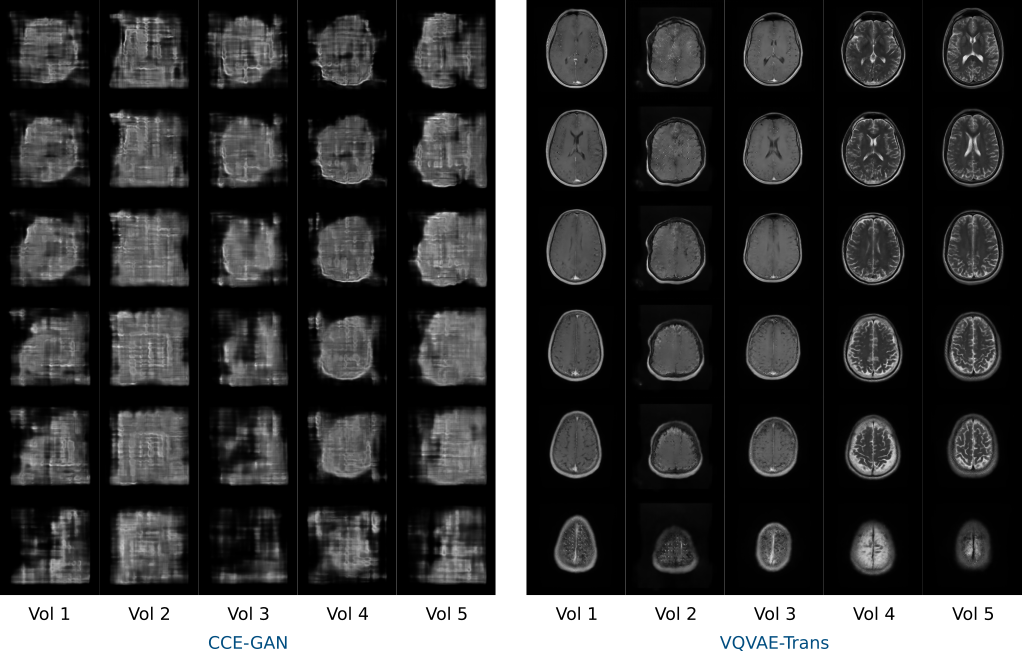}
\caption{Representative volumes synthesized by CCE-GAN and VQVAE-Trans in the fastMRI dataset. CCE-GAN is unable to synthesize realistic images.} 
\label{suppfig:nondiff_fastmri}
\end{figure}

\begin{figure}[h]
\centering
\includegraphics[width=1\linewidth]{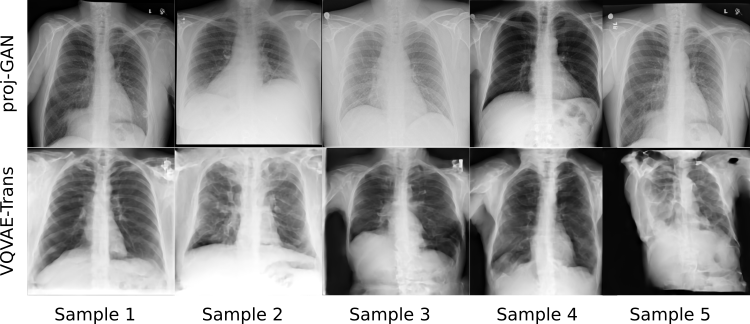}
\caption{Representative image synthesized by proj-GAN and VQVAE-Trans in the X-ray dataset. VQVAE-Trans is unable to synthesize high-quality images.} 
\label{suppfig:nondiff_xray}
\end{figure}

\textcolor{black}{We first assessed the prevalence of memorization in LDMs and compared it with other generative models. Afterward, we compared the quality of the synthesized images among all competing methods.
\subsubsection{Memorization in Latent Diffusion Models}}\label{results:memorization_ldms}
Theoretically, a model that perfectly learns the data distribution always has a non-zero probability of generating patient data copies.
As a result, a synthetic dataset with an infinite number of generated samples would eventually include all training samples.
Therefore, the critical question is determining how frequently the model generates patient data copies.
To answer this question, we synthesized a finite number of samples ($N_{syn}$) by setting it equal to the training data size ($N_{train}$). 
LDMs trained on each dataset were used to synthesize novel images. 
Afterwards, self-supervised models were used to detect potential replicas of training samples among the synthesized samples.
These self-supervised models first project the training, validation, and synthetic data onto a lower dimensional embedding space. This projection makes the copy detection process computationally efficient and further enables us to detect copies that are not just identical to the training samples but are also variations of the real samples such as flipping, rotation, and slight changes in contrast.
Next, Pearson's correlation coefficient reflecting similarity was computed between all pairs of training-validation and training-synthetic embeddings. Afterward, for each training embedding the closest validation and synthetic embedding were selected.
\begin{figure*}[t]
\centering
\includegraphics[width=0.8\linewidth]{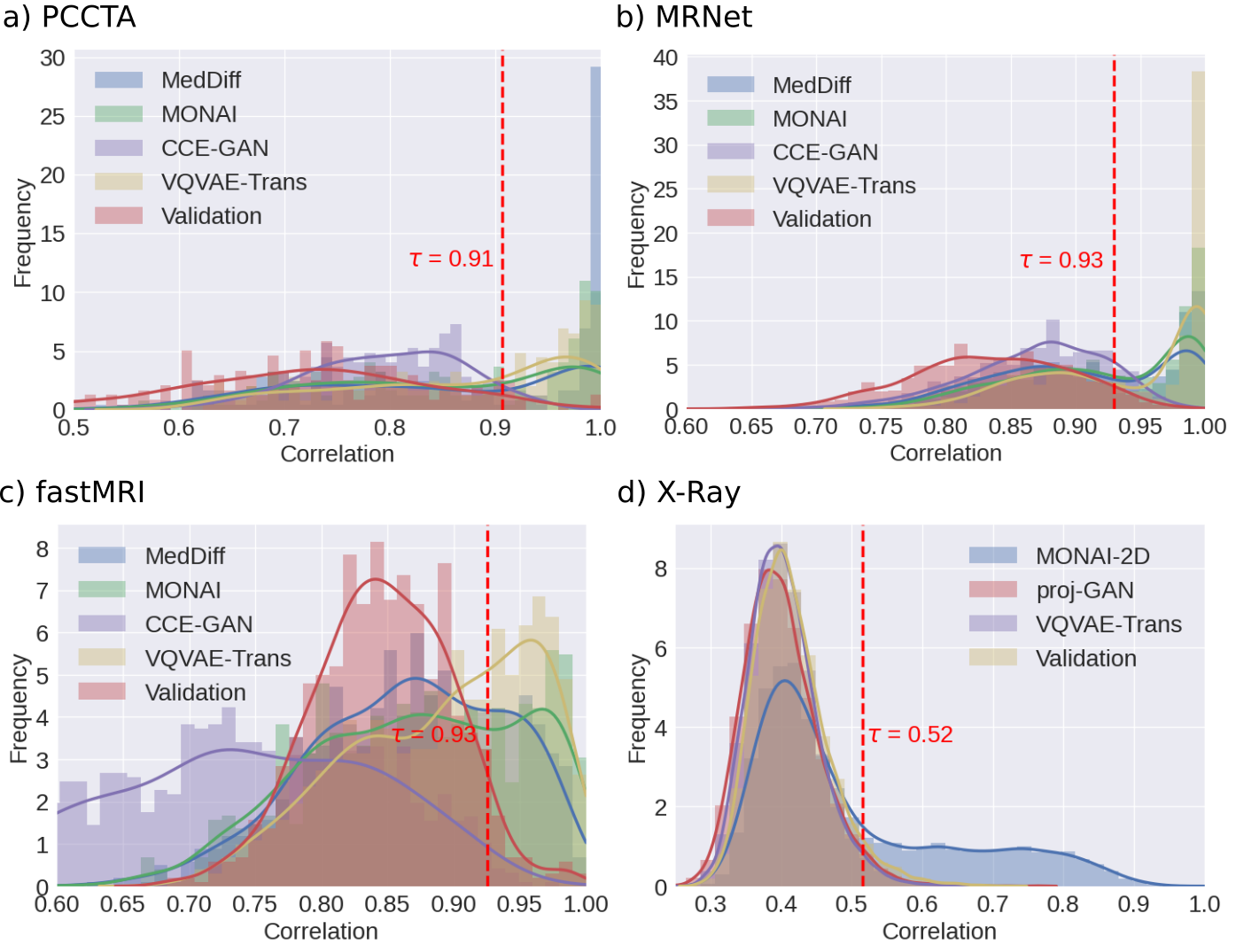}
\caption{Histograms showing distributions of Pearson's correlation values among closest training-validation pairs and training-synthetic pairs in a) PCCTA, b) MRNet, \textcolor{black}{c) fastMRI} and d) X-ray datasets. All training, validation, and synthetic samples were projected onto embedding space using self-supervised models. For each training embedding, closest embedding was selected from the validation data denoted as 'Validation' and \textcolor{black}{from the synthetic data for each generative model denoted as 'MedDiff', 'MONAI', 'MONAI-2D', 'CCE-GAN', 'proj-GAN' and 'VQVAE-Trans'}. Afterwards, $\tau$ was selected based on the 95th percentile of the correlation values in 'Validation' in each dataset, and synthetic samples with correlation values greater than $\tau$ were classified as copies.} 
\label{fig:histograms}
\end{figure*}
\begin{figure*}[h]
\centering
\includegraphics[width=1\linewidth]{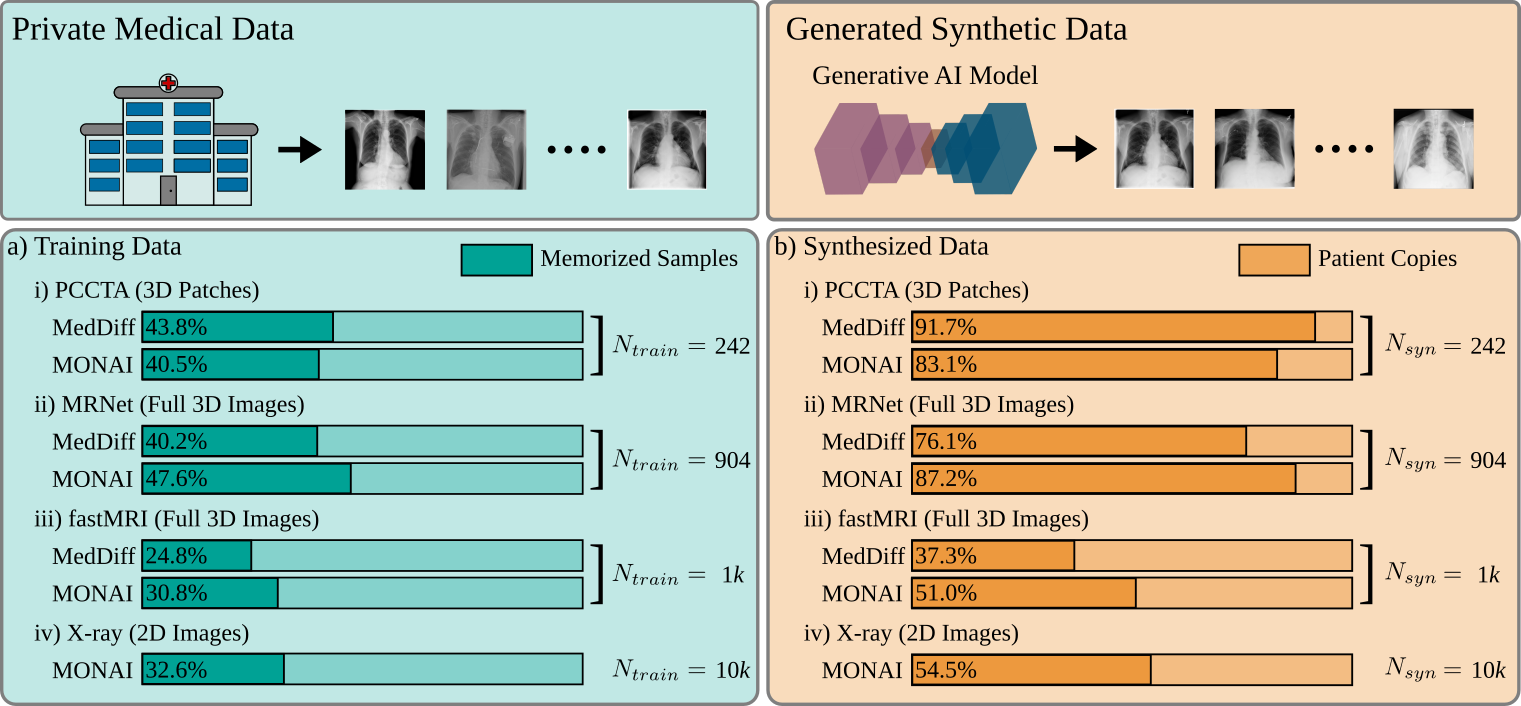}
\caption{ The left column represents private training data, and the right column represents synthesized data. a) Number of memorized training samples ($N_{mem}$) and b) number of synthesized samples that are patient data copies ($N_{copies}$) in PCCTA, MRNet, \textcolor{black}{fastMRI} and X-ray datasets as detected by our copy detection pipeline (Section. \ref{methods:copy_detection}). All datasets show a high percentage of $N_{mem}$ and $N_{copies}$, notably in 3D datasets. 
} 
\label{fig:mem_datasets}
\end{figure*}
\begin{figure}[h!b]
\centering
\includegraphics[width=1\textwidth]{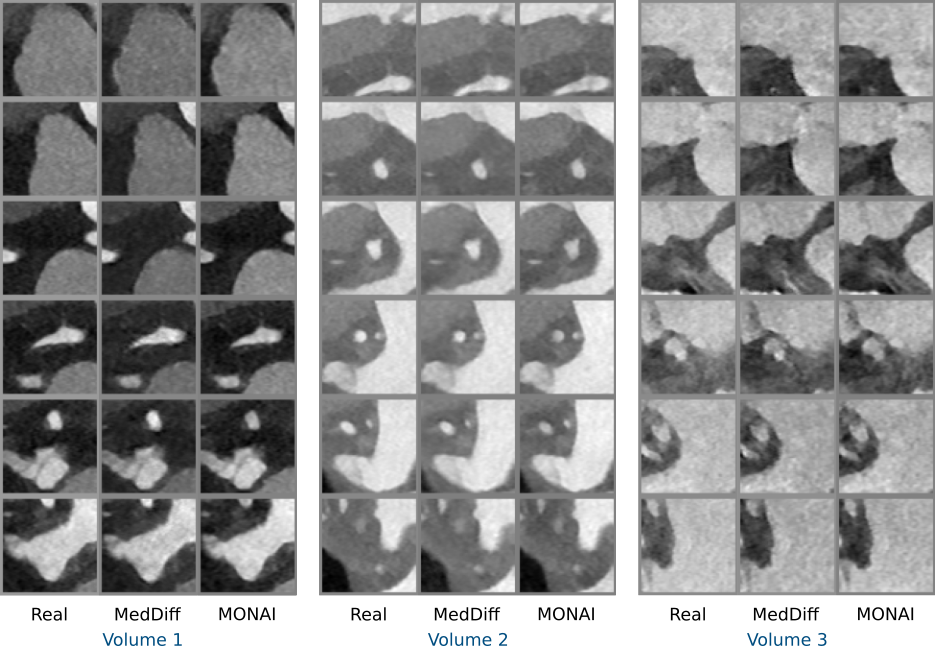}
\caption{Representative cross sections of real (Real) and copies (MedDiff, MONAI) detected in the PCCTA dataset. Copies show a high resemblance to the corresponding real samples across all slices.} 
\label{fig:3d_slices_pccta}
\end{figure}
Fig. \ref{fig:histograms} shows the distribution of correlation values between training and nearest validation embeddings ($\rho_{NN-val}$), and between training and nearest synthetic embeddings ($\rho_{NN-syn}$) in both MedDiff and MONAI. 
In all datasets and models, $\rho_{NN-syn}$ values were shifted more towards the right compared to $\rho_{NN-val}$ values, implying that synthetic samples bear a higher resemblance to the training data.\\
Next, we quantified the number of training samples memorized ($N_{mem}$) by the model and the number of synthesized samples that were copies ($N_{copies}$) based on a correlation threshold value $\tau$ (for details please refer to Section. \ref{methods:copy_detection}). The numbers are reported in Fig. \ref{fig:mem_datasets}.
In PCCTA dataset, (43.8, 40.5) \% of the training data were memorized in (MedDiff, MONAI),  and (91.7, 83.1) \% of the synthetically generated samples were identified as patient data copies in (MedDiff, MONAI). Fig. \ref{fig:3d_slices_pccta} shows copies that were detected in both MedDiff and MONAI along with the closest training samples. In the PCCTA dataset, which contains low-dimensional 3D patches, most of the details were preserved in the memorized samples, and in terms of quality, both MedDiff and MONAI generated images of similar quality.\\
\begin{figure}[t]
\centering
\includegraphics[width=1.0\textwidth]{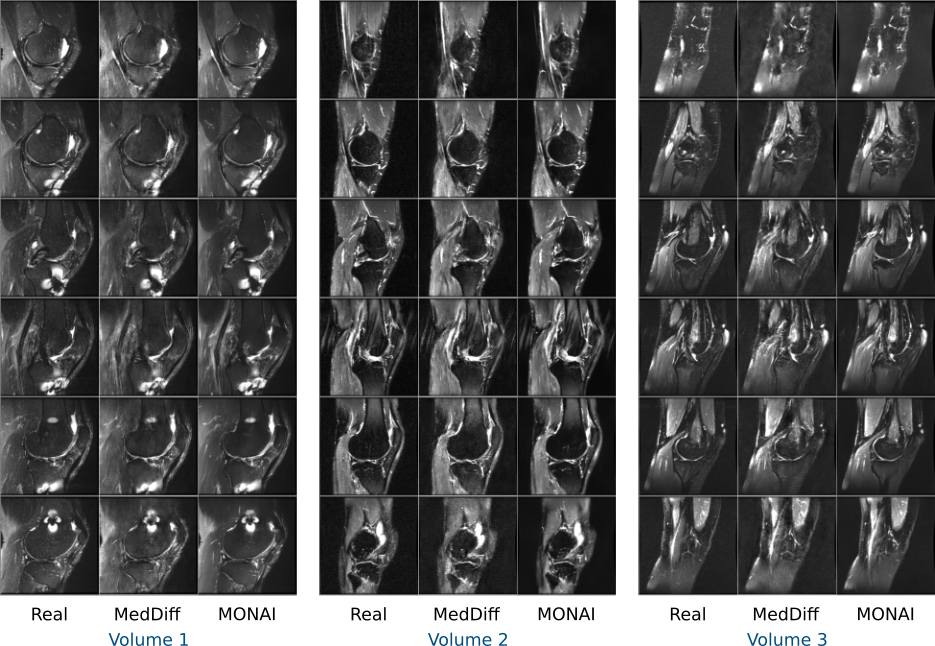}
\caption{Representative cross sections of real (Real) and copies (MedDiff, MONAI) detected in the MRNet datasets. Copies show a high resemblance to the corresponding real samples.} 
\label{fig:3d_slices_mrnet}
\end{figure}
\begin{figure}
\centering
\includegraphics[width=1.0\textwidth]{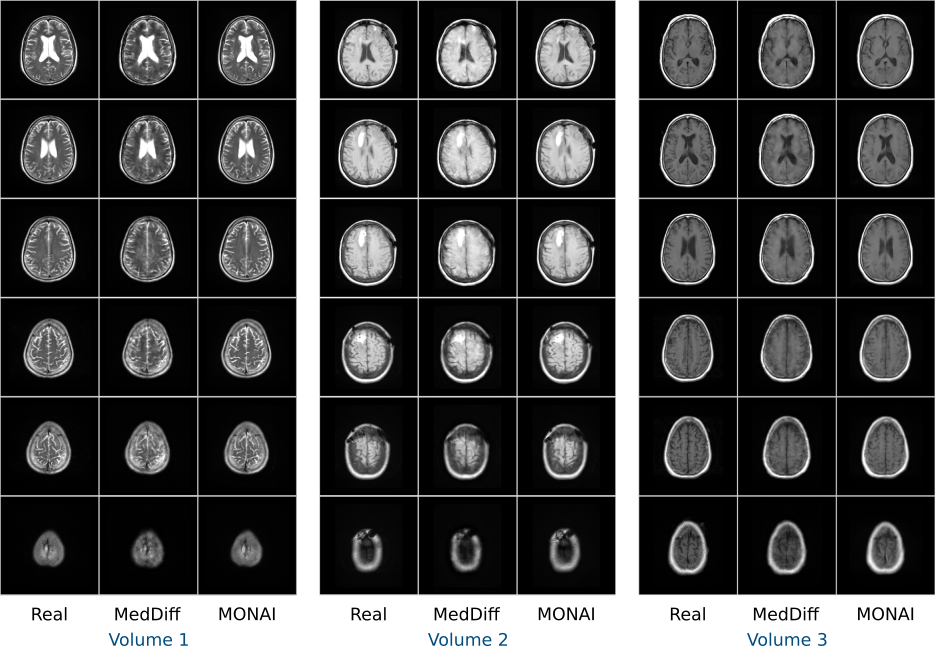}
\caption{\textcolor{black}{Representative cross sections of real (Real) and copies (MedDiff, MONAI) detected in the fastMRI datasets. Copies show a high resemblance to the corresponding real samples. MedDiff-synthesized images contain severe artifacts and only share global structures with the corresponding real images.}} 
\label{fig:3d_slices_fastmri}
\end{figure}
\begin{figure}[h]
\centering
\includegraphics[width=0.9\linewidth]{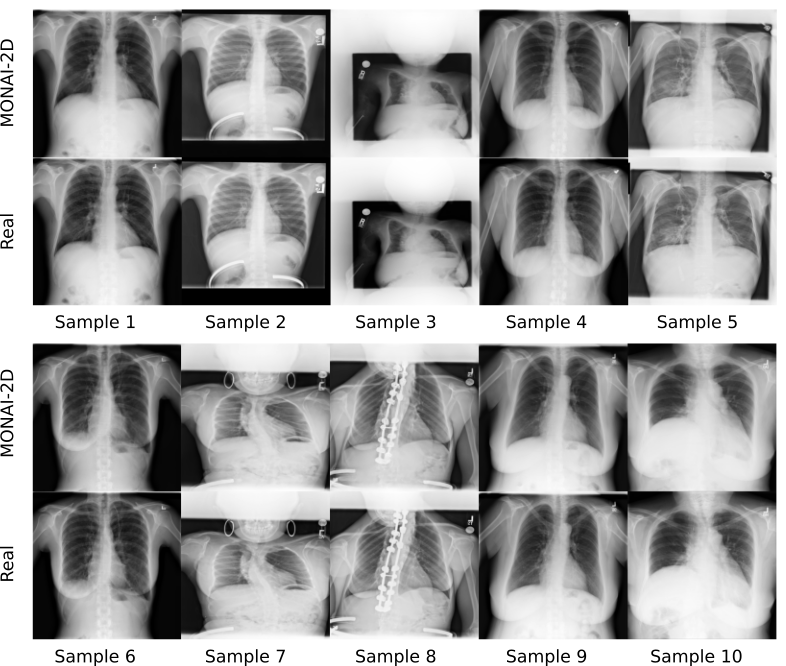}
\caption{Representative cross sections of real (Real) and copies (MONAI-2D) detected in the X-ray dataset. Copy candidates show a high resemblance to the corresponding real samples. The network tends to copy even the exact position of the image in case of partial field-of-view coverage.} 
\label{fig:nihxray_slices}
\end{figure}
\begin{figure}[h]
\centering
\includegraphics[width=1\linewidth]{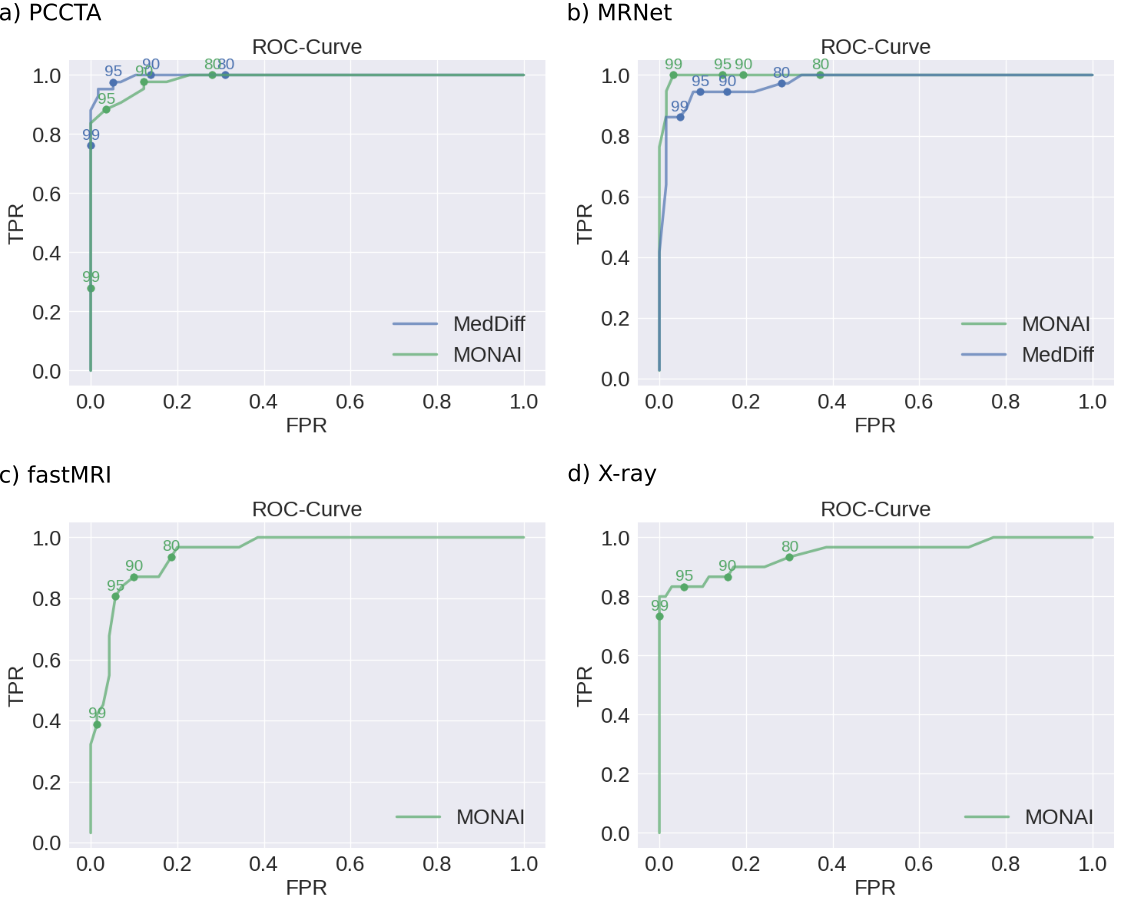}
\caption{\textcolor{black}{We randomly selected 100 training and nearest synthetic sample pairs, and a user manually labeled the corresponding nearest synthesized samples as novel or copies. These labels were then compared with the ones obtained based on $\tau$. Afterwards, false positive rates (FPRs) and true positive rates (TPRs) were calculated to obtain ROC curves. Threshold values based on (80, 90, 95, 99)th-percentile values are also marked.}} 
\label{fig:rocs}
\end{figure}
In MRNet dataset, (40.2, 48.2) \% of the training data were memorized in (MedDiff, MONAI),  and (76.1, 87.4) \% of the synthetically generated samples were identified as patient data copies in (MedDiff, MONAI). Fig. \ref{fig:3d_slices_mrnet} shows copies that were detected in both MedDiff and MONAI along with the closest training samples. In the MRNet dataset, which contained full 3D volumes, most of the global structure was preserved, albeit with notable differences in fine structural details between MedDiff and MONAI. MedDiff-synthesized images were noisy and unable to capture low-level structural details (Supp. Fig. \ref{suppfig:mrnet_slices}). MONAI-synthesized images, on the other hand, had a lower noise level but were slightly blurry (Supp. Fig. \ref{suppfig:mrnet_slices}). In MRNet, while both networks produced patient data copies, both models were unable to generate small structural details.\\
\textcolor{black}{In fastMRI dataset, (24.8, 30.8) \% of the training data were memorized in (MedDiff, MONAI),  and (37.3, 51.0) \% of the synthetically generated samples were identified as patient data copies in (MedDiff, MONAI). Fig. \ref{fig:3d_slices_fastmri} shows copies that were detected in both MedDiff and MONAI along with the closest training samples. MedDiff-synthesized images contained severe artifacts and were unable to capture low-level structural details. MONAI-synthesized images, on the other hand, had a lower noise level but were slightly blurry.\\}
In the 2D X-ray dataset, 32.6 \% of the training data were memorized and 54.5\% of the synthetic samples were patient data copies in MONAI-2D. Fig. \ref{fig:nihxray_slices} shows copies alongside the closest training samples. Synthetic samples show a very close resemblance to the training samples.
Overall, we observed a high level of patient data was memorized in both 2D and 3D models. Moreover, a very large percentage of synthesized samples were patient data copies, especially in 3D models. 
\textcolor{black}{\subsubsection{Memorization in Other Generative Models}\label{results:mem_gans_ae} 
We also assessed the phenomenon of memorization in some of the other widely used generative models. For this purpose, we evaluated the percentage of training data memorized and the amount of synthetic samples that are patient data copies in GANs and Autoencoders.
In 3D datasets, we included CCE-GAN \cite{ccegan} and VQVAE-Trans \cite{vqvaetransnmi}. In VQVAE-Trans an autoregressive generative model is trained in the latent space of a VQVAE. Supp. Tab. \ref{supptab:comparison_methods} lists the percentage of training data memorized and of synthetic samples that are patient data copies in all competing methods.
In VQVAE-Transformers (49.6, 58.3, 40.2)\% of the training data are memorized in (PCCTA, MRNet, fastMRI) and (66.1, 83.3, 57.5)\% of the synthesized samples are detected as copies (PCCTA, MRNet, fastMRI). Our copy detection model also detected some of the samples synthesized by CCE-GAN as copies. However, CCE-GAN was unable to synthesize realistic samples (Supp. Figs. \ref{suppfig:nondiff_pccta}, \ref{suppfig:nondiff_mrnet}, and \ref{suppfig:nondiff_fastmri}) and most of the detected copies only shared very little global information with the training samples. In 2D datasets, we included proj-GAN \cite{projgan} and VQVAE-Transformers. Here VQVAE-Transformers could not synthesize images of high quality (Supp. Fig. \ref{suppfig:nondiff_xray}) and therefore there were no copies among the synthesized samples (only a very small number of samples falsely detected as copies). Proj-GAN synthesized images of reasonable quality (Supp. Fig. \ref{suppfig:nondiff_xray}) and the synthesized images contain no copies (only a very small number of samples falsely detected as copies). 
Overall, LDMs consistently suffer from higher levels of memorization across both 2D and 3D datasets. However, LDMs have the ability to synthesize better images overall. To further confirm this expert radiologists graded images synthesized by all competing methods. In the 3D datasets, 10 samples were randomly selected from real training images and (MONAI, MedDiff, CCE-GAN, VQVAE-Trans) synthesized images, making 50 images in total. Afterward, their order was shuffled and they were presented to expert radiologists to grade the images based on the criteria mentioned in section \ref{methods:rad_eval_crit}.
Supp. Figs. \ref{suppfig:rad1} and \ref{suppfig:rad2} show scores as graded by the radiologists. Overall, MONAI was the best in terms of all criteria except for realism in MRNet where VQVAE-Trans was better, and anatomical correctness in (MRNet, fastMRI) where one expert reports similar scores. 
The same experiments were repeated in the 2D X-ray dataset, where 20 samples were randomly selected from real training images and (MONAI-2D, proj-GAN, VQVAE-Trans) synthesized images, making 60 images in total. Afterward, they were randomly shuffled and presented to the radiologists. 
Here, again, MONAI-2D outperformed all methods. In 2D synthesis, VQVAE-Trans was unable to synthesize images of reasonable quality. Proj-GAN on the other hand was close to MONAI-2D. Taken together, the expert evaluation from the 2D and 3D datasets suggests that the LDM-based model,i.e., MONAI outperforms their counterpart generative models. }
\subsection{Efficient Copy Detection}\label{results:copy_detection}
\begin{figure}[h]
\centering
\includegraphics[width=1\linewidth]{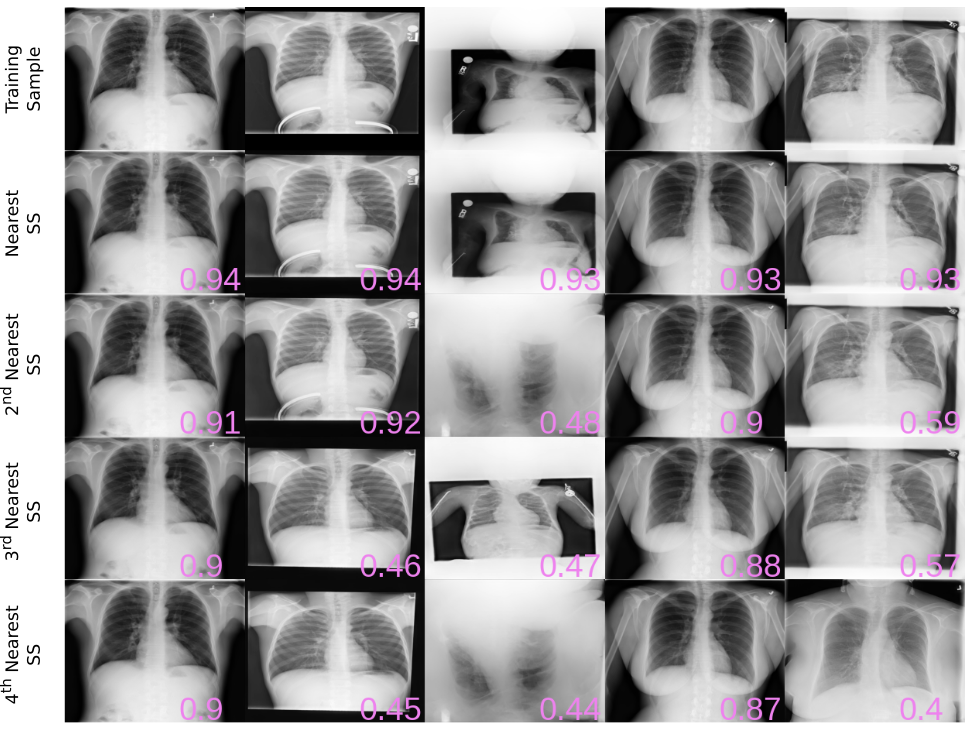}
\caption{Representative training samples are shown along with their nearest, 2\textsuperscript{nd} nearest, 3\textsuperscript{rd} nearest, and 4\textsuperscript{th} nearest synthetic samples (SSs) in the embedding space of the self-supervised method are shown. Correlation values of the SSs with the corresponding training samples are also shown in violet color. The correlation values appear to be good indicators of the samples' similarity. The correlation threshold ($\tau$) for copy detection in this experiment was 0.52. Note that all SSs having $\tau$ greater than 0.52 appear to be patient data copies. } 
\label{suppfig:train_nn}
\end{figure}

\begin{figure}[h]
\centering
\includegraphics[width=1\linewidth]{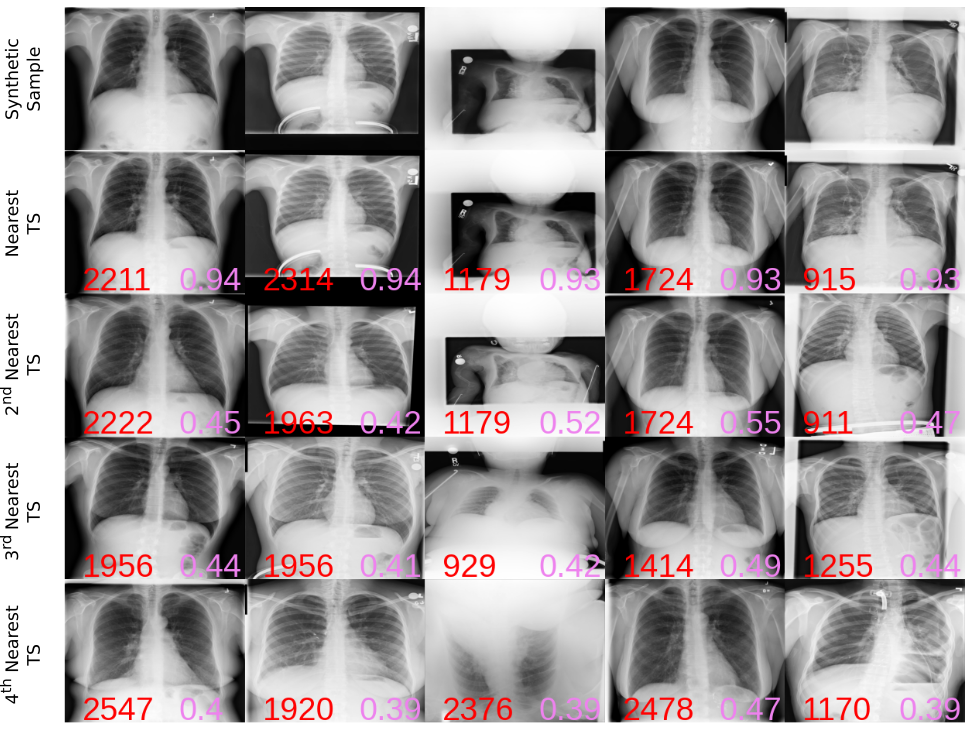}
\caption{Representative synthetic samples are shown along with their nearest, 2\textsuperscript{nd} nearest, 3\textsuperscript{rd} nearest, and 4\textsuperscript{th} nearest training samples (TSs) in the embedding space of the self-supervised method are shown. Patient numbers of the TSs are shown in red color, and correlation values of the training samples TSs with the corresponding synthetic samples are also shown in violet color. The correlation values appear to be good indicators of the samples' similarity. The correlation threshold ($\tau$) for copy detection in this experiment was 0.52, and appears to be a good indicator of patient copy detection.} 
\label{suppfig:synth_nn}
\end{figure}

\begin{figure}[h]
\centering
\includegraphics[width=1\linewidth]{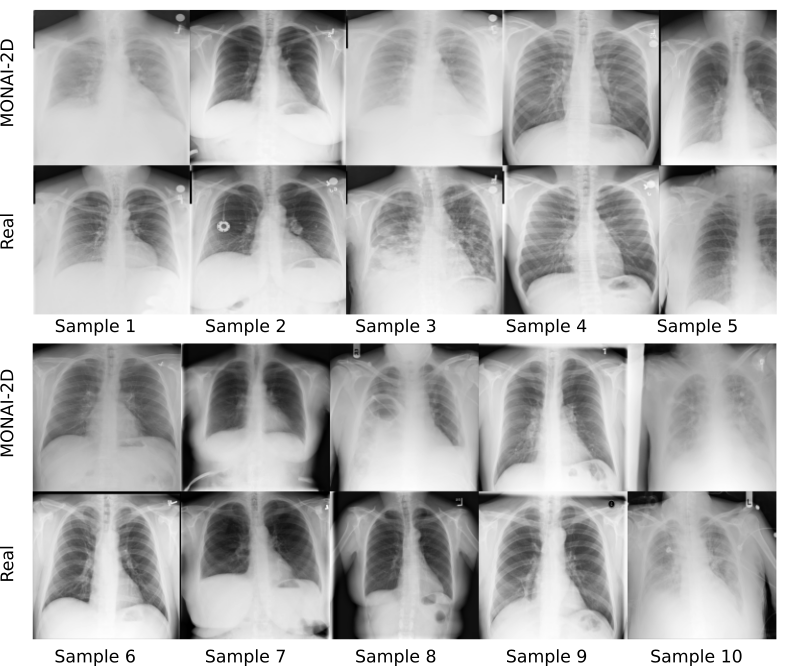}
\caption{Representative copy candidates falsely classified as patient data copies are shown along with their corresponding real training samples. Copy candidates show similarity at a global level with the training samples.} 
\label{suppfig:synth_nn_dummy}
\end{figure}

\begin{figure}[h]
\centering
\includegraphics[width=0.7\linewidth]{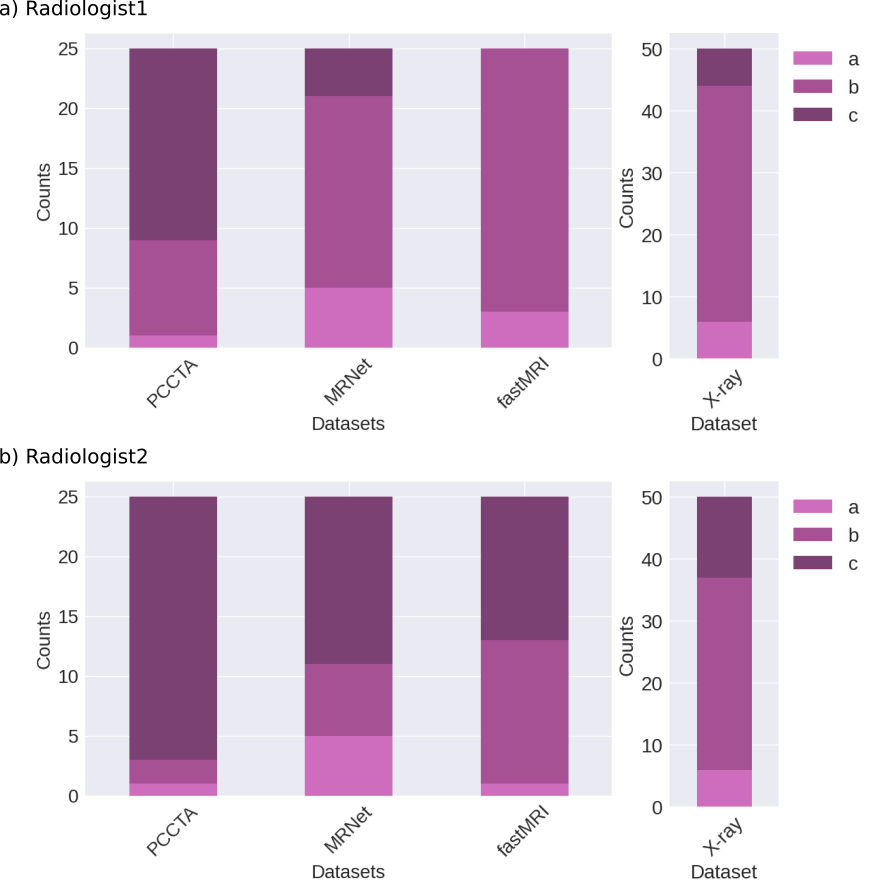}
\caption{Expert evaluation scores on detected copies in PCCTA, MRNet, fastMRI, and X-ray datasets. Two experts evaluated the detected copies by manually scoring them as either (a) not a copy (b) a copy with minor structural variations such as vessels, plaques, foreign materials, etc., or (c) an exact copy with minor variations such as rotation, flipping, and change in contrast/brightness} 
\label{suppfig:rad_score_cd}
\end{figure}

\begin{figure}[h]
\centering
\includegraphics[width=0.8\linewidth]{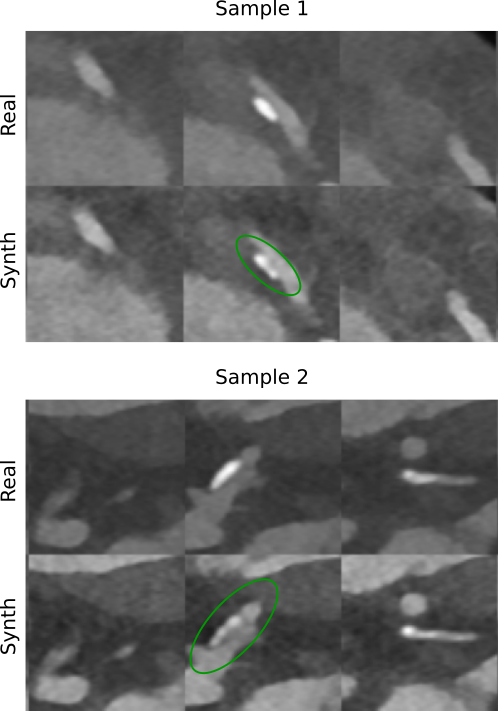}
\caption{Detected copies categorized as copies with minor structural variations by an expert radiologist in the PCCTA dataset are shown. Three cross-sections of two training samples (Real) and the corresponding detected copies (Synth) showing structural variations were selected. The calcified plaque in the synthesized samples is different from the corresponding training samples.} 
\label{suppfig:rad_cd_pccta}
\end{figure}

\begin{figure}[h]
\centering
\includegraphics[width=0.8\linewidth]{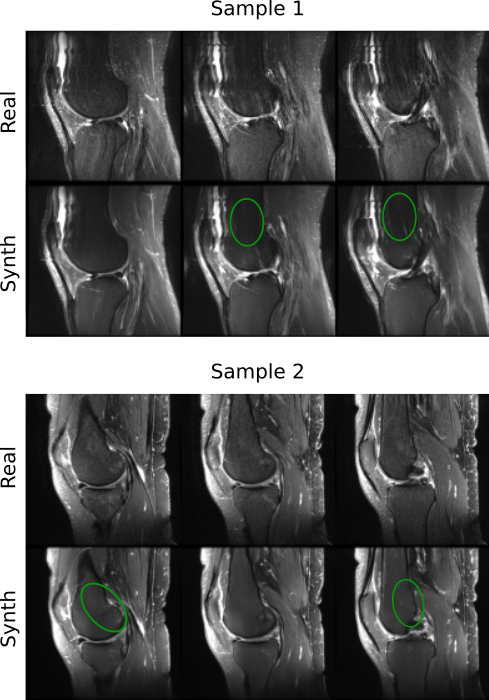}
\caption{Detected copies categorized as copies with minor structural variations by an expert radiologist in the MRNet dataset are shown. Three cross-sections of two training samples (Real) and the corresponding detected copies (Synth) showing structural variations were selected. The artifacts (Sample 1) and bone marrow signal (Sample 2) in the synthesized samples are different from the corresponding training samples.} 
\label{suppfig:rad_cd_mrnet}
\end{figure}

\begin{figure}[h]
\centering
\includegraphics[width=0.8\linewidth]{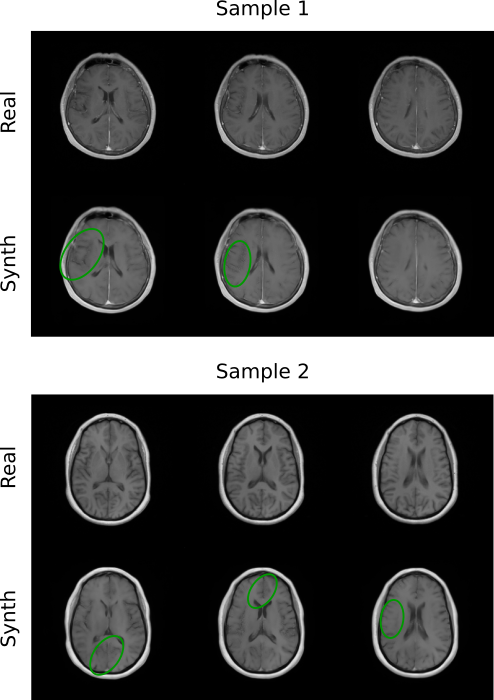}
\caption{Detected copies categorized as copies with minor structural variations by an expert radiologist in the fastMRI dataset are shown. Three cross-sections of two training samples (Real) and the corresponding detected copies (Synth) showing structural variations were selected. Vessels are mostly missing from the synthesized Sample 1, and the gray-white matter transitions in the synthesized Sample 2 are blurred.} 
\label{suppfig:rad_cd_fastmri}
\end{figure}

\begin{figure}[h]
\centering
\includegraphics[width=0.9\linewidth]{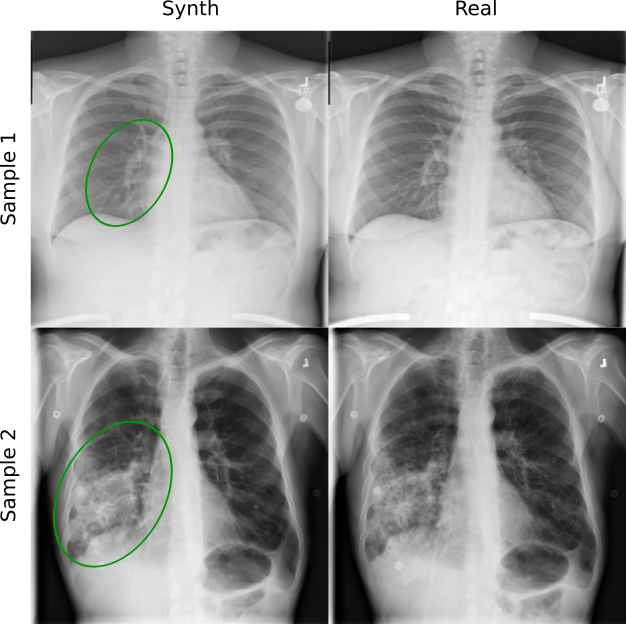}
\caption{Detected copies categorized as copies with minor structural variations by an expert radiologist in the X-ray dataset are shown. Two training samples (Real) and the corresponding detected copies (Synth) showing structural variations were selected. Vessels are missing in synthesized Sample 1 and the pathology is more homogenous in synthesized Sample 2.} 
\label{suppfig:rad_cd_xray}
\end{figure}

Next, we gauged the effectiveness of the self-supervised models in detecting copies. 
\textcolor{black}{\subsubsection{Performace}\label{results:copy_detection_perf}
We first assessed the performance of the copy detection models.}
The correlation threshold $\tau$ used to categorize samples as copy or novel was based on the 95th percentile of the correlation values between training and nearest validation sample embeddings (Fig. \ref{fig:histograms}). 
To evaluate if the selected threshold value $\tau$ was meaningful, we also randomly selected 100 training and nearest synthetic sample pairs, and a user manually labeled the corresponding nearest synthesized samples as novel or copies. These labels were then compared with the ones obtained based on $\tau$. 
In the PCCTA dataset, copies were detected with sensitivity of \textcolor{black}{97.6\%} and specificity of \textcolor{black}{93.1\%} among the MedDiff-synthesized samples and with sensitivity of \textcolor{black}{88.4\%} and specificity of \textcolor{black}{94.7\%} among MONAI-synthesized samples.
In the MRNet dataset, copies were detected with sensitivity of \textcolor{black}{95.4\%} and specificity of \textcolor{black}{90.6\%} in the MedDiff-synthesized samples and with sensitivity of \textcolor{black}{100\%} and specificity of \textcolor{black}{85.5\%} in the MONAI-synthesized samples.
\textcolor{black}{In the fastMRI dataset, copies were detected with sensitivity of 80.6\% and specificity of 94.3\% in the MONAI-synthesized samples. MedDiff-synthesized samples contained severe artifacts, and were left out of this analysis.}
In the X-ray dataset, copies were detected with sensitivity of \textcolor{black}{83.3\%} and specificity of \textcolor{black}{94.2\%} in the MONAI-2D synthesized samples.
These values reflect the effectiveness of the copy detection pipeline. 
Supp. Fig. \ref{suppfig:train_nn} also shows training samples alongside their four closest synthetic samples in the embedding space of the self-supervised models and Supp. Fig. \ref{suppfig:synth_nn} shows synthetic samples alongside their four closest training samples in the embedding space of the self-supervised models. Correlation values in the embedding space and the chosen $\tau$ values appear to be reliable indicators for detecting patient data copies.\\
\textcolor{black}{\subsubsection{Robustness}\label{results:copy_det_reobust}
The copy detection approach in this study relies on the self-supervised model trained to attract a sample to its variations and repel a sample from other samples in the lower-dimensional embedding space. The quality of the lower-dimensional embedding space and hence the performance of the self-supervised model can be affected by several factors. 
\paragraph{Batch Size:} One important factor is the batch size which defines the number of positive and negative pairs within a batch (Supp. Eq. \ref{suppeqn:ntxent}). In our training setting based on the Normalized Temperature-scaled Cross Entropy (NT-XENT) loss function, each iteration contains n-1 positive pairs and n*(n-1) negative pairs, where n is the batch size. A large batch size means that the loss is weighed more for the negative pairs, and a small batch size means that the loss function is weighed similarly for positive and negative pairs, which can bring all samples close to each other and  impact the model's ability to differentiate different-looking samples from similar samples more efficiently. This can potentially reduce the true positive rate of the self-supervised method.
To test the effect of batch size on the performance of the copy detection, models were trained for batch size in \{2, 5, 10 (baseline), 20\} in the 3D datasets and \{32, 64 (baseline), 128\} in the 2D dataset. 
Supp. Tab. \ref{supptab:comparison_batch_size} lists \% $N_{mem}$ and $N_{copy}$ in all datasets.
In all cases, except for PCCTA, $N_{mem}$ and $N_{copy}$ were low at small batch sizes (2 in 3D and 32 in 2D), and similar for medium and large batch sizes. Since the baseline 3D and 2D models had a very high true positive rate, this backs our hypothesis that a small batch can reduce the true positive rate. In the PCCTA dataset, we did not observe such trends. This can be attributed to the very small size of the dataset that could at times produce noisy results. 
\paragraph{Embedding Size:} Another important factor is the size of the lower dimensional embedding. A low dimension can limit the expressivity of the model and a very high dimension can make the model overfit to the variations only within the training data. 
To investigate the effect of lower dimensional embedding size, we trained contrastive learning models in \{16, 32 (baseline), 64\} in the 3D dataset and \{32, 64, 128 (baseline), 256\} in the 2D dataset. 
Supp. Tab. \ref{supptab:comparison_embedding_size} lists \% $N_{mem}$ and $N_{copy}$ in all datasets.
We observed changing dimensions did not have a significant impact on both $N_{mem}$ and $N_{copies}$. This suggests that the self-supervised models show little sensitivity to embedding dimensions.
\paragraph{Threshold:} 
The selection of threshold value ($\tau$) depends on the risk deemed acceptable while classifying a copy candidate as copy or novel. Depending on the requirements, this threshold value can be adjusted to the desired value. A threshold value higher than the optimum value can decrease the true positive rate, while a lower value can increase the false positive rate. 
In our study, the threshold value ($\tau$) was selected based on the 95th percentile of the correlation values between embeddings of the training samples and the closest validation sample. 
The rationale behind using this approach is that the 95th percentile of the distribution here corresponds to the upper tail of the distribution and any sample having a higher correlation would have a very high resemblance to the training data. 
To assess how the selection of threshold value affects the copy detection performance, we also assessed copy detection performance at different threshold values. Fig. \ref{fig:rocs} shows the Receiver operating characteristic (ROC) curve at different threshold levels. We observed that increasing the threshold decreased the false positive rate but also decreased the true positive rate, and decreasing the threshold increased both false positive rate and true positive rates. ROC curves across all datasets show that the copy detection approach is able to perform reasonably well at threshold values based on a range of percentile values.
Furthermore, $\tau$ selected based on the 95th percentile lies within the range where the true positive rate is high and the false positive rate is low.
\paragraph{Number of Synthetic Samples ($N_{synth}$):}Increasing the number of synthesized samples can increase $N_{mem}$ and $N_{copy}$ (Supp. Tab. \ref{supptab:comparison_synthetic_samples_rates}). But it could also increase the false positive rate. While detecting copy candidates for each training sample, each training sample is compared to a pool of all synthetic samples, the closest sample is assigned as a copy candidate if the correlation value between the training samples and the copy candidate is greater than $\tau$ it is assigned as a copy. However, as $N_{synth}$ increases we would expect the probability of a copy candidate being falsely categorized as a copy to also increase because although two images might not be identical, they can show very high similarity levels especially if the pool from which synthetic images are taken is very large. To understand this phenomenon, we performed memorization assessment on data samples that were not used during training using $N_{synth}$ = (5k, 10k, 20k). Ideally, we would expect the model to not detect any of the training samples memorized. 
Supp. Tab. \ref{supptab:comparison_synthetic_samples_rates_dummy} lists \% $N_{mem}$ and $N_{copy}$ in all datasets.
In fastMRI we observed percentage $N_{mem}$ = (1.2, 2.2, 4.6) for $N_{synth}$ = (0.5k, 1k, 2k), and in X-ray we observed percentage $N_{mem}$ = (3.8, 5.9, 9.5) for $N_{synth}$ = (5k, 10k, 20k). This suggests that although increasing $N_{synth}$ increases both $N_{mem}$ and $N_{copy}$, it also increases the probability of falsely classifying a copy candidate as a copy. Supp. Fig. \ref{suppfig:synth_nn_dummy} shows some of the copy candidates falsely classified as copies. Many of the copy candidates show a resemblance at a global level with the corresponding training samples. 
}
\textcolor{black}{\subsection{Quality of the Detected Copies}\label{results:rad_eval_copies} 
Our copy detection approach was able to detect copies among the synthesized samples, and the efficiency of the approach was determined via manual labeling of training samples and their corresponding copy candidates. To assess how close the detected copy candidates are compared to their corresponding training samples, two experts also evaluated the detected copies by manually scoring them as either (a) not a copy (b) a copy with minor structural variations such as vessels, plaques, foreign materials, etc. or (c) an exact copy with minor variations such as rotation, flipping, and change in contrast/brightness. The expert evaluation was performed on the best-performing overall method in 2D and 3D datasets, i.e., MONAI and MONAI-2D.  In (Xray, PCCTA, MRNet, fastMRI) datasets, (50, 25, 25, 25) randomly selected copies were graded. 
Supp. Fig. \ref{suppfig:rad_score_cd} shows the scores as graded by the radiologists.
In the PCCTA dataset, expert 1 classified (1, 8, 16) samples and expert 2 classified (1, 2, 22) samples under classes (a, b, c). Notable differences between synthesized samples under class b and their corresponding real samples were differences in terms of calcified plaques (Supp. Fig. \ref{suppfig:rad_cd_pccta}).
In the MRNET dataset, expert 1 classified (5, 16, 4) samples and expert 2 classified (5, 6, 14) samples under classes (a, b, c). Notable differences between synthesized samples under class b and their corresponding real samples differences in terms of small bone structures and missing artifacts (Supp. Fig. \ref{suppfig:rad_cd_mrnet}).
In the fastMRI dataset, expert 1 classified (3, 22, 0) samples and expert 2 classified (1, 12, 12) samples under classes (a, b, c). Notable differences between synthesized samples under class b and their corresponding real samples were missing vessels, missing basal ganglia, and gray-white matter transitions (Supp. Fig. \ref{suppfig:rad_cd_fastmri}).
In the X-ray dataset, expert 1 classified (6, 38, 6) samples and expert 2 classified (6, 31, 13) samples under classes (a, b, c). Notable differences between synthesized samples under class b and their corresponding real samples were missing small vessels and foreign objects (Supp. Fig. \ref{suppfig:rad_cd_xray}).
We believe that these differences can be attributed to the ability of the model to synthesize samples with very sharp and small-scale details. While the models synthesize realistic images, they are unable to capture very small structures within the images. 
}
\subsection{Factors Affecting Memorization}
Next, we investigated the factors that could potentially influence memorization \textcolor{black}{in LDMs}. For this purpose, we considered three different aspects including training data size, training iterations, data augmentation \textcolor{black}{and network architecture size. To reduce computational complexity, unless stated, the experiments were performed on the best-performing method as evaluated by the experts, i.e., MONAI in 3D- and MONAI-2D in 2D-datasets.  
}
\subsubsection{Impact of Training Data Size}\label{results:train_data_size}
Deep neural networks are prone to overfitting upon training on small datasets. Although overfitting and memorization are distinct concepts, overfitting can lead to memorization.
LDMs learn data generation through gradual denoising, an inherently ill-posed problem with infinitely many solutions. Training LDMs on small datasets makes the models overfit to solutions, leading to denoised training images. This can in turn increase the likelihood of generating more training samples at random.
To explore this phenomenon, we investigated the effect of training data size ($N_{train}$) on memorization in the \textcolor{black}{fastMRI and }X-ray dataset\textcolor{black}{s}, the main reason being that \textcolor{black}{these are large datasets} and provides us with the freedom to select different numbers of training samples.
\textcolor{black}{In the fastMRI dataset, we compared the LDMs trained for $N_{train}$ = (1k, 2k, 5k) images and in the X-ray dataset, we compared the LDMs trained for $N_{train}$ = (5k, 10k, 20k) images. We considered two different scenarios. In the first scenario, we maintained the number of epochs at (2k, 3k) for all three models in (fastMRI, X-ray) to ensure that each model encountered each training sample the same number of times. The self-supervised model was trained on (5k, 20k) training images in (fastMRI, X-ray).}
\textcolor{black}{Supp. Tab. \ref{supptab:comparison_Ntrain_epochs}} shows the percentage of training samples that were memorized and the percentage of synthesized samples that were patient data copies as detected by the copy detection pipeline.
Increasing the training set size reduced the percentage of memorized training data samples. However, surprisingly, there was a slight increase in the number of memorized training samples ($N_{mem}$) with the increase in the number of training samples ($N_{train}$), \textcolor{black}{with $N_{mem}$ equal to (143, 255, 368) for ($G_{\theta, 1k}$, $G_{\theta, 2k}$, $G_{\theta, 5k}$) in fastMRI} and equal to (2.7k, 3.3k, 3.5k) for ($G_{\theta, 5k}$, $G_{\theta, 10k}$, $G_{\theta, 20k}$) in Xray. 
\textcolor{black}{The number of synthesized samples that were patient data copies ($N_{copies}$) did not show a consistent correlation for both datasets. This can explained by the fact that ideally $N_{copies}$ should always be higher than $N_{mem}$, and although the percentage $N_{mem}$ is decreasing, the number $N_{mem}$ is increasing. So depending on the number $N_{mem}$, $N_{copies}$ might increase or decrease by increasing $N_{train}$. Another observation is that for fstMRI at $N_{train}$=5k, $N_{copies}$ is higher than $N_{mem}$. This can be attributed to one synthetic sample falsely categorized as a copy candidate of more than one training sample.}\\
\textcolor{black}{In the second scenario, we kept the number of iterations the same as oftentimes the generative models are trained for a fixed number of iterations. So we trained models for a fixed number of iterations equal to (150k, 100k) in (fastMRI, X-ray). 
Supp. Tab. \ref{supptab:comparison_Ntrain_itr} shows the percentage of training samples that were memorized and the percentage of synthesized samples that were patient data copies as detected by the copy detection pipeline.
We observed that in such cases, in the fastMRI dataset, the models trained on 5k images ($G_{\theta, 5k}$) memorize very little data as opposed to when the number of epochs was fixed (4.2 vs 7.4)\%. In the X-ray dataset, we observed similar results. The models trained on 20k images ($G_{\theta, 20k}$) memorize very little data as opposed to when the number of epochs was fixed (5.7 vs 17.7)\%. The experiments conducted under both scenarios suggest that the ability of a sample to be memorized depends on both the number of times a sample is seen during training and the number of training samples.}

\begin{figure}[h!]
\centering
\includegraphics[width=1\linewidth]{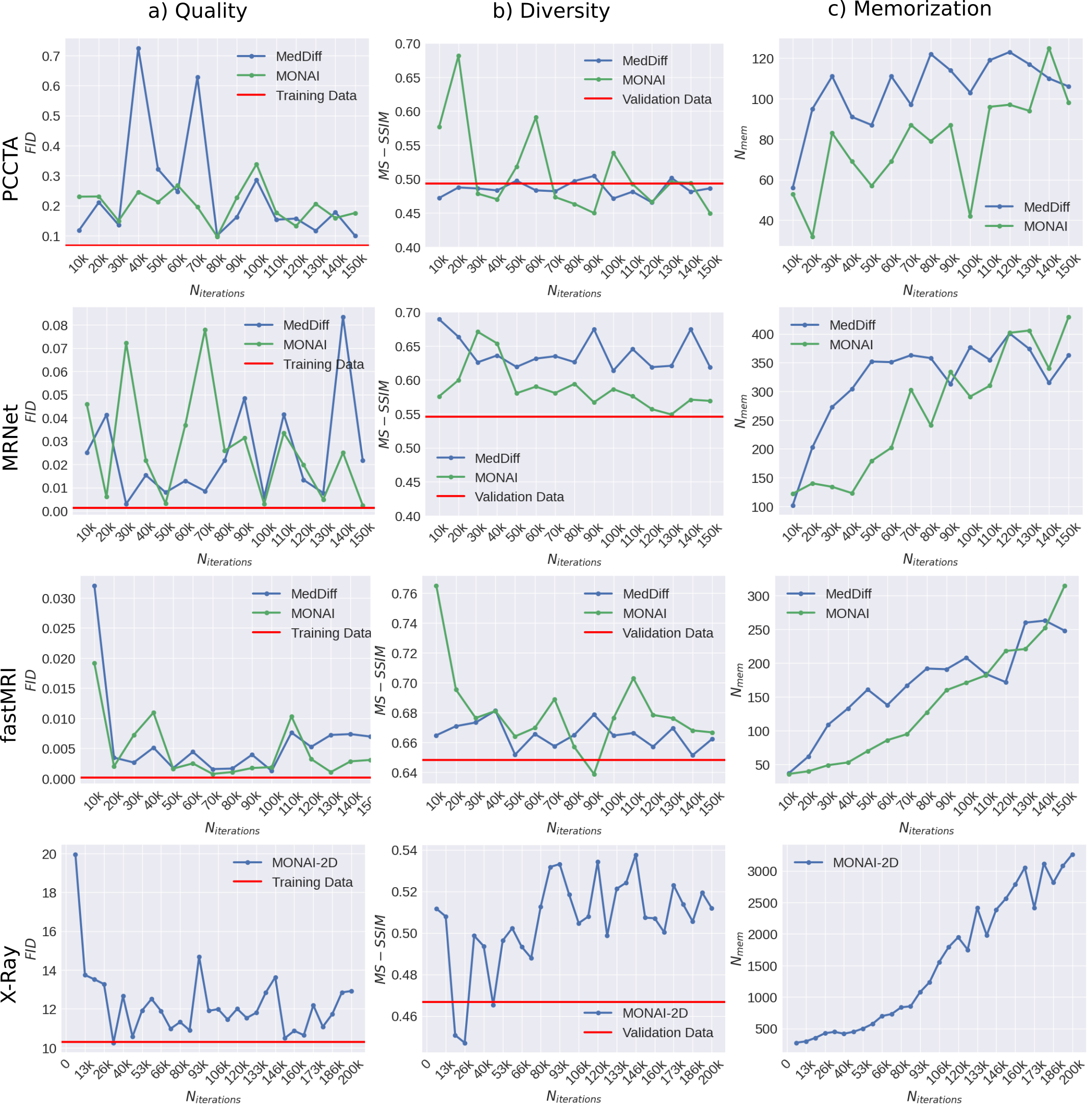}
\caption{a) Quality assessment metric FID between synthesized (MedDiff, MONAI, MONAI-2D) and real validation images, and between real training (Training Data) and real validation images are shown. b) Diversity assessment metric MS-SSIM within the synthesized data (MedDiff, MONAI, MONAI-2D), and within the validation data (Validation Data) are shown. c) Number of memorized training samples ($N_{mem}$) among the synthesized samples (MedDiff, MONAI, MONAI-2D) as a function of number of iterations ($N_{iterations}$) used for training are shown.} 
\label{fig:epochs}
\end{figure}
\subsubsection{Memorization as a Metric}\label{results:iterations}
One aspect of LDMs that receives little attention is the number of iterations or epochs used for training, and most of the studies just report a number without performing a thorough evaluation. Overtraining the network can make the network overfit to the training data while denoising, and can lead to more frequent generation of training samples during progressive denoising. This can lead to enhanced memorization \cite{dar2024epoch}.
To investigate the effect of training epochs/iterations on memorization of LDMs, we calculated the number of memorized training samples ($N_{mem}$) as a function of training iterations.
Fig. \ref{fig:epochs}c shows $N_{mem}$ detected as a function of training iterations ($N_{iterations}$). 
In all datasets, $N_{mem}$ increased with training iterations, suggesting that over-training the model can lead to enhanced memorization. \\
In addition to the relation between $N_{mem}$ and $N_{iterations}$, we were also interested in evaluating their relation with metrics conventionally used for assessment or training of generative models.
For this purpose, we also calculated Fréchet inception distance (FID) \cite{FID} which measures the quality of the synthesized samples, and multi-scale structural similarity index measure (MS-SSIM) \cite{MS-SSIM} which quantifies diversity among the synthesized samples as a function of $N_{iterations}$. Lower FID suggests high quality and low MS-SSIM indicates high diversity.
Ideally, we expect FID to decrease and then converge at a point. However, in \textcolor{black}{PCCTA and fastMRI} datasets, FID did not follow a fixed pattern for both models (Fig. \ref{fig:epochs}a) and showed large variations across $N_{iterations}$. This is alarming since FID is perhaps one of the most widely used metrics to assess image quality and compare data-generating capabilities with other models.
\textcolor{black}{In the fastMRI dataset, FID first decreased and started oscillating, and both MedDiff and MONAI show similar values of FID for most of the iterations. While these curves seem meaningful, the radiological evaluation in section \ref{results:mem_gans_ae} showed that MONAI-synthesized images were much higher quality than fastMRI.}
In the 2D X-ray dataset, FID decreased till 33k iterations and then oscillated afterward. This suggests that image quality saturates after a specific number of iterations. 
Across all datasets, \textcolor{black}{with the exception of MONAI in fastMRI}, MS-SSIM did not display a consistent trend (Fig. \ref{fig:epochs}b). Ideally, MS-SSIM should be low, indicating high diversity.
However, one thing to consider is that MS-SSIM only quantifies diversity and does not provide any information regarding image quality. In fact, a model generating random noise can have very high diversity.
Taken together, our results suggest that traditional measures used to quantify the quality and diversity of the synthesized samples can be misleading and memorization is an aspect that should be taken into account while training generative models, and perhaps a hybrid metric could be used for training models for open-data sharing \cite{Borji2022metrics}.
\begin{figure*}[h]
\centering
\includegraphics[width=1\linewidth]{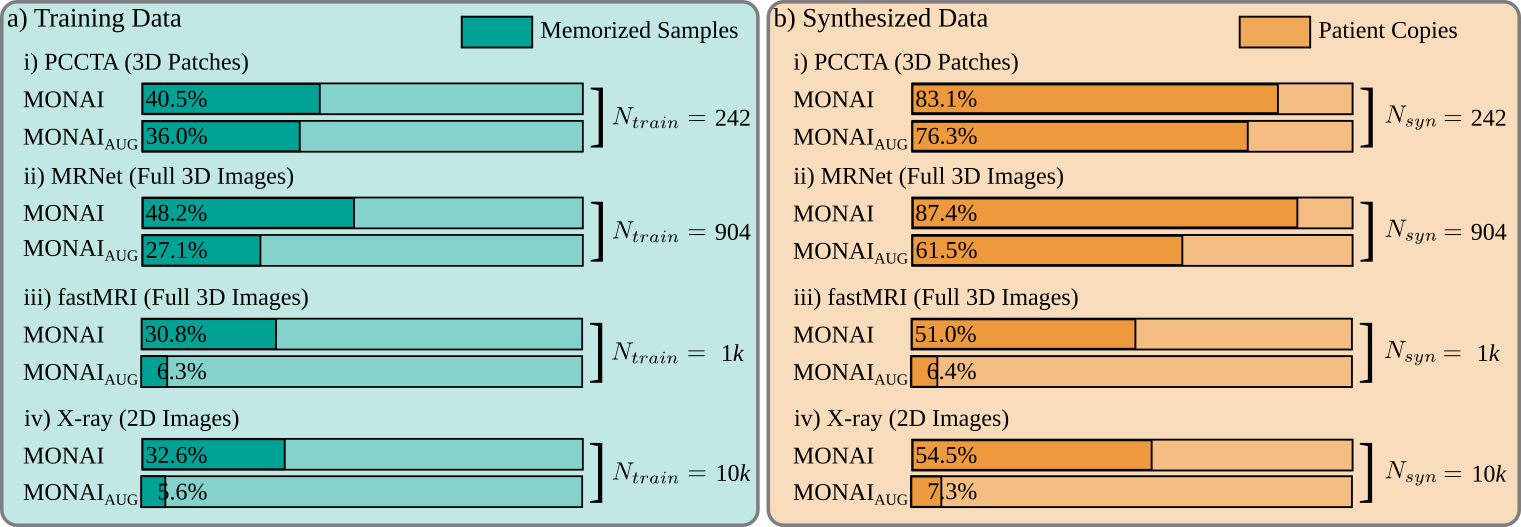}
\caption{ Percentage a) $N_{mem}$ and b) $N_{copies}$ are shown in the MONAI models trained with (MONAI\textsubscript{AUG}) and without (MONAI) data augmentation. Augmentation during training reduces patient data memorization. 
} 
\label{fig:mem_aug_monai}
\end{figure*}
\subsubsection{Mitigation via Data Augmentation}\label{results:augmentation}
\begin{figure}[h]
\centering
\includegraphics[width=1\linewidth]{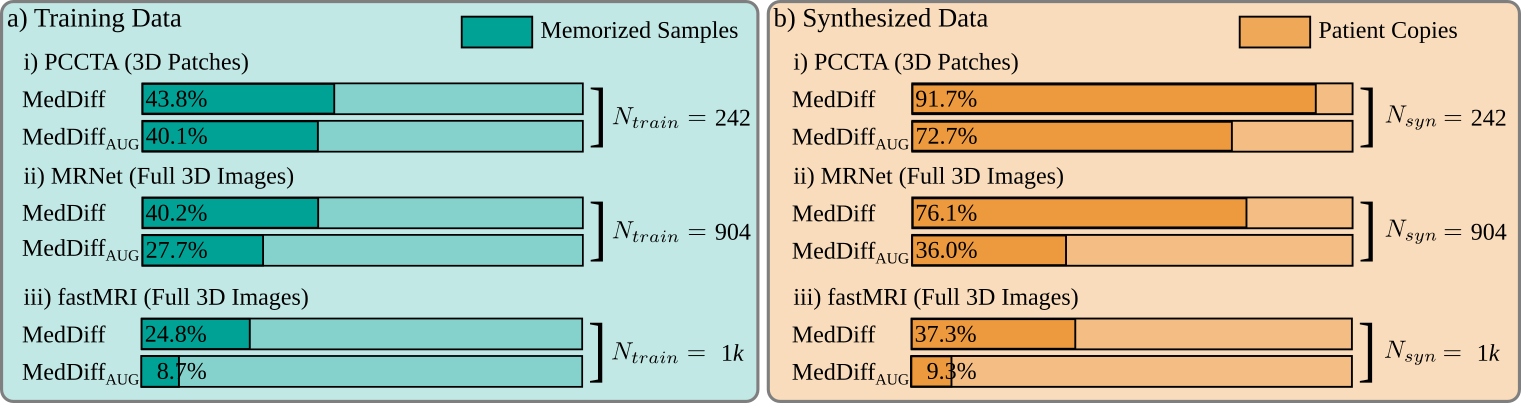}
\caption{ Percentage a) $N_{mem}$ and b) $N_{copies}$ are shown in MedDiff models trained with (MedDiff\textsubscript{AUG}) and without (MedDiff) data augmentation. Augmentation during training reduces patient data memorization.
} 
\label{suppfig:mem_aug_meddiff}
\end{figure}

\begin{figure}[h]
\centering
\includegraphics[width=1\linewidth]{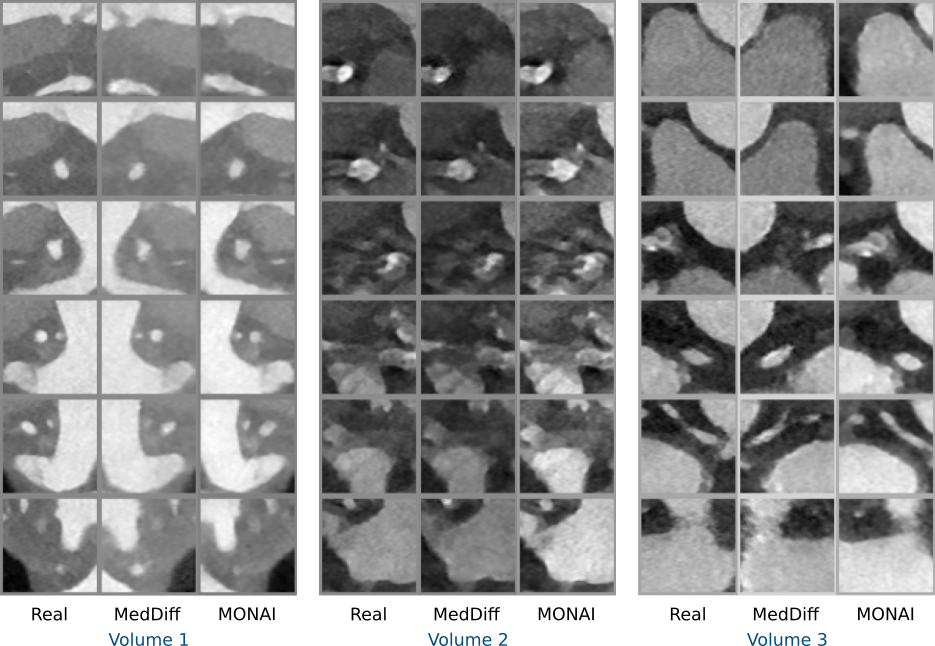}
\caption{Representative cross sections of real (Real) and copies (MedDiff, MONAI) synthesized by augmented models in the PCCTA dataset.} 
\label{suppfig:3d_slices_pccta_aug}
\end{figure}

\begin{figure}[h]
\centering
\includegraphics[width=1.0\linewidth]{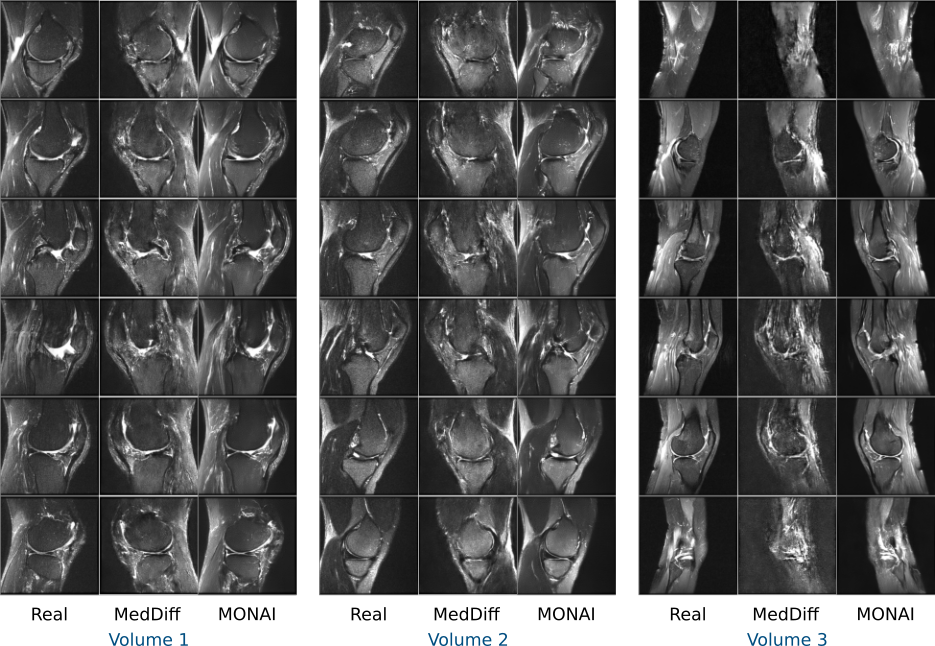}
\caption{Representative cross sections of real (Real) and copies (MedDiff, MONAI) synthesized by augmented models in the MRNet dataset. } 
\label{suppfig:3d_slices_mrnet_aug}
\end{figure}

\begin{figure}[h]
\centering
\includegraphics[width=1.0\linewidth]{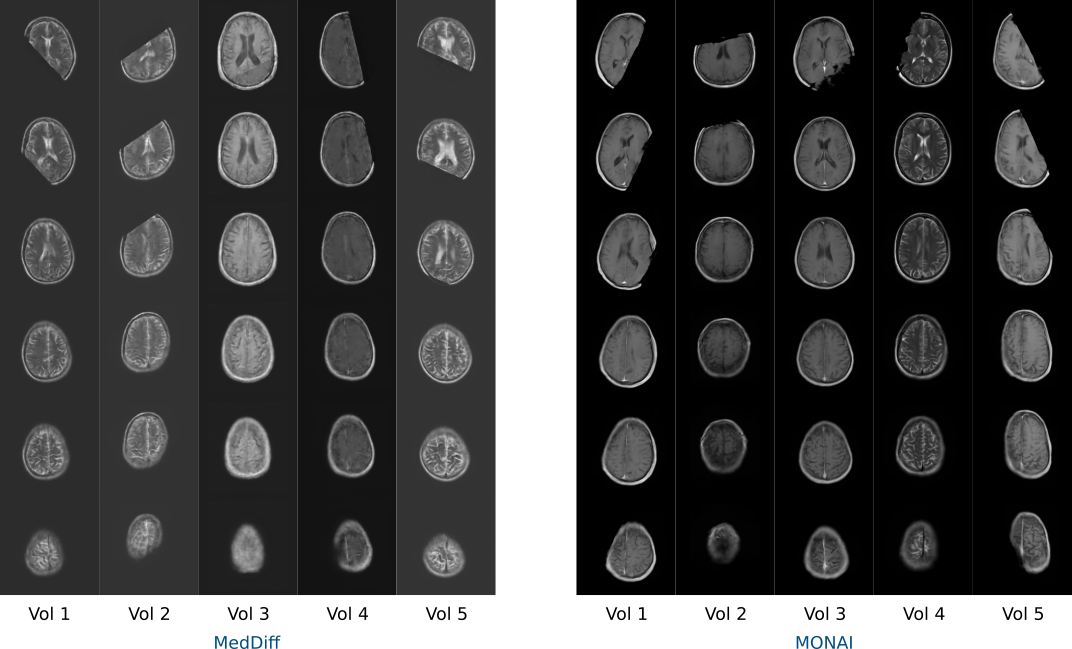}
\caption{Representative cross sections of volumes synthesized by MedDiff\textsubscript{AUG} and MONAI\textsubscript{AUG} in the fastMRI dataset. Not carefully performing data augmentation during training can lead to the synthesis of unrealistic images.} 
\label{suppfig:3d_slices_fastmri_aug}
\end{figure}

\begin{figure}[h]
\centering
\includegraphics[width=1\linewidth]{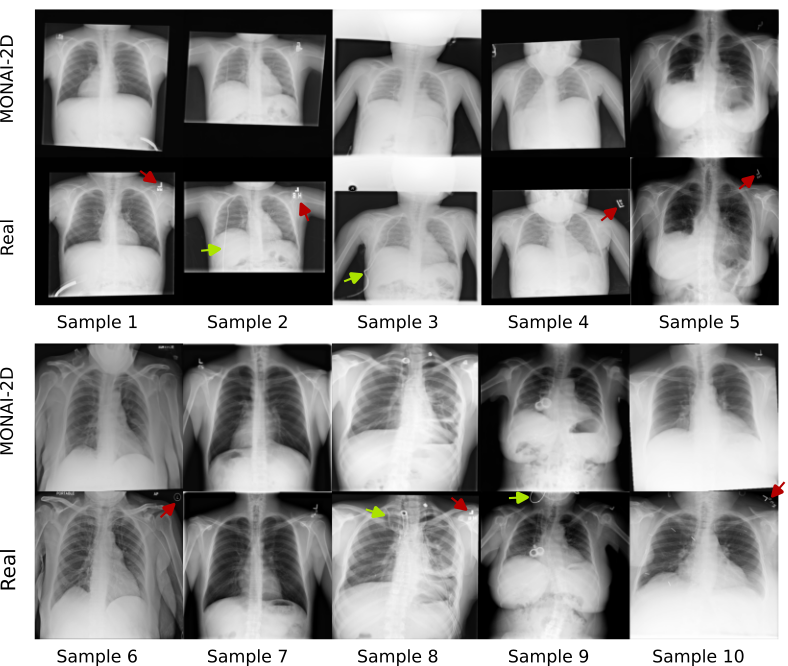}
\caption{Representative cross sections of real (Real) and copies (MONAI-2D) detected in the X-ray dataset. The augmented model generates copies that in addition to augmentations also produce minor variations (marked with arrows).} 
\label{suppfig:nihxray_slices_aug}
\end{figure}

Data augmentation is a widely used technique that artificially expands the training dataset size by complementing it with variations of training samples. This typically enhances generalizability in deep neural networks, potentially reducing memorization in LDMs \cite{dar2023investigating}. Here, we also assessed memorization in LDMs by training models on expanded datasets obtained via augmentation techniques (MedDiff\textsubscript{Aug}, MONAI\textsubscript{Aug}). 
In each epoch, all training samples underwent flipping and rotation (between $-5^{\circ}$ to $5^{\circ}$ along all axes) operations with a probability of 50\% each. Fig. \ref{fig:mem_aug_monai} \textcolor{black}{and Supp. Fig. \ref{suppfig:mem_aug_meddiff}} compare percentage $N_{mem}$ and $N_{copies}$ in \textcolor{black}{MONAI and MedDiff} models trained with and without data augmentation as detected by the copy detection pipeline.\\
In the PCCTA dataset, (40.1, 36.0) \% of the training data was memorized in (MedDiff\textsubscript{Aug}, MONAI\textsubscript{Aug}),  and (72.7, 76.3) \% of the synthetically generated samples were identified as patient data copies in (MedDiff\textsubscript{Aug}, MONAI\textsubscript{Aug}). This suggests a decrease in memorization in both models. Supp. Fig. \ref{suppfig:3d_slices_pccta_aug} shows copies that were detected in both MedDiff\textsubscript{Aug} and MONAI\textsubscript{Aug} along with the closest training samples. The copy detection approach is able to capture copies that are also variations of the training samples. In volume 1, the synthesized samples are flipped versions of the training samples, and in volume 3, MedDiff\textsubscript{Aug} synthesized sample is flipped and MONAI\textsubscript{Aug} synthesized sample is a rotated version of the training sample.  \\
In the MRNet dataset, (27.7, 27.1) \% of the training data was memorized in (MedDiff\textsubscript{Aug}, MONAI\textsubscript{Aug}),  and (36.0, 61.5) \% of the synthetically generated samples were identified as patient data copies in (MedDiff\textsubscript{Aug}, MONAI\textsubscript{Aug}). We also observed a decrease in memorization in the MRNet dataset. Supp. Fig. \ref{suppfig:3d_slices_mrnet_aug} shows copies that were detected in both MedDiff\textsubscript{Aug} and MONAI\textsubscript{Aug} along with the closest training samples. The copy detection pipeline was able to detect copies. In MRNet, the samples generated via MedDiff\textsubscript{Aug} were very poor (Supp. Fig. \ref{suppfig:mrnet_slices}). Although they resemble the corresponding training images globally, they are unable to generate high-quality images. MONAI\textsubscript{Aug} was able to retain the image quality, however, similar to the models trained without augmentation, we observed slight blurriness induced in the synthesized images.\\
\textcolor{black}{In the fastMRI dataset, (8.7, 6.3) \% of the training data was memorized in (MedDiff\textsubscript{Aug}, MONAI\textsubscript{Aug}),  and (9.3, 6.4) \% of the synthetically generated samples were identified as patient data copies in (MedDiff\textsubscript{Aug}, MONAI\textsubscript{Aug}). We observed a sharp decline in memorization in the fastMRI dataset. This decrease is significantly more than PCCTA and MRNet datasets. We believe that this sharp decline could be explained by the unrealistic images produced by augmentation. Since the images already contain very few axial slices, learning from augmented data can lead to the synthesis of unrealistic images (Supp. Fig. \ref{suppfig:3d_slices_fastmri_aug}), which can also effect the performance of the generative models.\\}
In the 2D X-ray dataset,  5.6\% of the training samples were memorized, and 7.3\% of the synthesized samples were copies. This suggests a drastic reduction in memorization compared to the 3D models, which had a significantly smaller training set.
Supp. Fig. \ref{suppfig:nihxray_slices_aug} shows some of the selected copies. A careful examination of the copies shows that the patient data copies are not just augmented versions of the original patient image. These copies also contain some notable minor structural variations.
One potential explanation is that such models generalize well as they come across different variations of the training samples and this artificial expansion gives them the ability to generate samples that are not identical to the training samples but interpolations of different variations of the same sample. 
Another explanation could be related to the way such models are trained. LDMs are trained to perform denoising. In the case of augmentation, the model comes across variations of each training sample multiple times, and instead of finding a solution that produces denoised training images, the model can converge to a solution that is based on minimizing training error across all variations. This in turn can produce images that are not identical to a training sample or its variations, but rather a solution that is an average of all variations. This can potentially lead to the removal or blurriness of small structures.
For instance, if we carefully observe samples 2, 3, 8, and 9 (Supp. Fig. \ref{suppfig:nihxray_slices_aug} green markers), we can see that a structure that resembles a wire is present in the training samples but missing from the copies. Furthermore, in samples 1, 2, 4, 5, 6, 8, and 10 (Supp. Fig. \ref{suppfig:nihxray_slices_aug} red markers), the character 'L' at the top right corner is blurred out in the copies.\\
One intriguing observation is that although we observed a decrease in memorization due to data augmentation during training in both 2D and 3D LDMs, the reduction in memorization was moderate in \textcolor{black}{PCCTA and MRNet} but huge in \textcolor{black}{fastMRI and X-ray}. Furthermore, the synthesized images in 2D-LDMs contained prominent notable variations compared to 3D-LDM. The underlying reason could be hard to speculate because of differences in dataset features such as training data size, anatomy, and resolution. 
Overall, we observed that using data augmentation reduced memorization. However, we observed that it could also lead to compromise in image quality. 
\textcolor{black}{\subsubsection{Impact of Network Architecture }\label{results:architecture}
While deep neural networks have demonstrated state-of-the-art performance in many medical imaging tasks, they are also prone to overfitting. A large network with few samples can be prone to memorizing patient-specific features instead notably in low data regimes. 
To understand how the number of parameters can affect memorization, a more fair comparison would be between models trained on the same dataset. To understand this phenomenon we trained models with (small, medium (baseline), large) architecture sizes using (25m, 171m, 270m) trainable parameters in MONAI (2D) and (68m, 191m, 442m) trainable parameters in MMONAI-3D, and assessed memorization in all models. Supp. Tab. \ref{supptab:comparison_architecture_size2}  shows $N_{mem}$ and $N_{copes}$ among all datasets in MONAI and MONAI-2D. Overall across all datasets, with the exception of PCCTA, we observed that the small architecture had a lower level of memorization, and medium and large architectures had a higher but similar level of memorization. The differences in PCCTA can be attributed to the small data size, which can be even sufficient for the small model to memorize. This suggests that network architectures should be carefully selected prior to training generative models and not carefully selecting model architecture can have an impact on memorization.}
\section{Discussion}
In this work, we assessed memorization in unconditional latent diffusion models for medical image synthesis. Trained models were used to synthesize novel medical images, and potential copies were detected using self-supervised models based on the contrastive learning approach. Our results obtained on different datasets covering various organs, resolutions, fields of view, contrasts, and modalities indicate that such models are prone to patient data memorization. Furthermore, our self-supervised models were able to identify copies among synthetic images with reasonable performance levels. 
Additional complementary analyses point out several factors that can have an impact on memorization. Adding data augmentation operations reduced memorization, whereas over-training enhanced memorization. Increasing the training data size slightly increased the number of memorized samples, however, it decreased the probability of a synthesized sample being a patient data copy. These results suggest that memorization could be mitigated to some extent with careful training.\\
\textcolor{black}{While memorization has been extensively studied in the field of computer vision \cite{wei2024memorizationdeeplearningsurvey,yoon2023diffusion,Carlini2023extracting,Somepalli2023,Somepalli2023understanding} and natural language \cite{wei2024memorizationdeeplearningsurvey,Carlini2021extracting}, its role and prevalence within the medical imaging community remain largely unexplored and insufficiently understood. Medical images introduce unique challenges, including limited data availability, variable resolutions, finer structural details, and distinct imaging characteristics. The extent of memorization in medical images is not yet well understood under these aforementioned settings. Moreover, the memorization phenomenon observed in computer vision has primarily been demonstrated on 2D images. In our study, we focus on 3D medical images, which present additional complexities. Thus, a more appropriate comparison would be between generative models trained on video data and those trained on 3D medical images. To our knowledge, no studies have addressed memorization in generative models for video data. Our findings indicate that latent diffusion models can also memorize 3D images.} 
\\
To date, only a handful of studies have investigated patient data memorization in medical imaging \cite{dar2023investigating, dar2024epoch,akbar2023beware}. Akbar et al. \cite{akbar2023beware} assessed memorization in 2D diffusion models, and observed higher pixel-wise correlation among synthetic and real training samples as opposed to real test and training samples. 
In our previous work \cite{dar2023investigating}, we conducted experiments on 3D imaging datasets and utilized contrastive learning to suggest patient data copy candidates for each training sample in a lower dimensional latent space. 
In another prior investigation,  we showed that overtraining can lead to enhanced memorization \cite{dar2024epoch}.
Compared to previous studies, we conducted a more comprehensive evaluation of memorization across various datasets. We assessed both memorized training samples and the synthetic samples that were patient data copies. Additionally, we proposed and analyzed an approach to detect them and investigated underlying reasons in different datasets, which could assist in alleviating memorization.\\
\textcolor{black}{The proposed copy-detection method effectively identifies patient data copies and demonstrates robustness to hyperparameter choices. However, we observed that increasing the number of synthesized samples leads to a higher false positive rate in detecting memorized training samples. This indicates that the approach is more suitable for smaller synthetic datasets. Nonetheless, it is worth noting that a higher false positive rate might be acceptable in this context, as detecting patient data copies, which pose a greater risk, remains the primary concern.} \\
It could be argued that memorization could be mitigated by simply tracking validation error and avoiding over-fitting. However, this assumes equivalency between over-fitting and memorization. While the two terms might sometimes be correlated, this categorization is inaccurate \cite{MemorizationVsOverfitting}. Over-fitting is a global phenomenon where models attain very high accuracy on the training data typically at the expense of test data accuracy. Memorization, on the other hand, corresponds to the assignment of a very high likelihood to the training data points. As a matter of fact, memorization of a model can be enhanced even when validation loss decreases, especially in the earlier phases of training when memorization might be increasing but the test loss might be decreasing \cite{MemorizationVsOverfitting}. \\
\textcolor{black}{In our study, we observed that LDMs tend to memorize a diverse set of medical images with varying properties, such as resolution, organ type, complexity, and modality. This suggests that all types of data have the potential to be memorized. However, understanding why certain data are more prone to memorization is challenging and can not be directly inferred from our study. For effective training of generative models, large and diverse datasets are typically used to help models learn the underlying data distribution. When a distribution has multiple modes (whatever the models might represent), LDMs can efficiently learn modes that are well-represented by many samples. However, modes with only a few samples may not be adequately learned; instead, the model might memorize these examples as discrete impulses that directly reflect the training data. This suggests that unique or rare samples in the dataset could be more susceptible to memorization. Nevertheless, we believe that it is an important aspect that definitely warrants future work.} \\
Our results indicate crucial factors for memorization. \textcolor{black}{We can design our experiments to mitigate memorization during both training and synthesis. Our results suggest that the higher the number of samples are seen by the latent diffusion models, the more likely the data is to be memorized. So, one straightforward strategy could be to reduce the number of times a certain image is seen by utilizing as much training data as possible \cite{akbar2023beware,akbar2024brain} and performing data augmentations during training.
Some studies also train 2D generative models on slices from 3D volumes \cite{diffusionanomaly,diffusion4medreview,diffusionrecon}. While this might increase the number of training samples for the 2D generative model, such models lose the global contextual information and are unable to work in settings where synthetic volumes are desired. 
Another way to reduce memorization would be to ensure that we do not over-train our models, and devise and utilize early stopping criteria. Memorization can also be mitigated through memorization-informed guided synthesis such that the embeddings of the synthesized samples are different from the embeddings of the training samples. In addition to performing memorization-informed synthesis, we can also perform memorization-informed training. This can be done by performing denoising during training to make the denoised image slightly different from the training data by not making it overfit to the training data while ensuring that the denoised image is also realistic. This can be performed through adversarial training with the help of a discriminator.} Fernandez et al. \cite{fernandez2023privacy} proposed a two-step approach \textcolor{black}{for privacy-preserved open-data sharing}. In the first step, a diffusion model was trained on real data and the synthesized samples were refined to contain only novel samples. These refined samples were then utilized to train a new model with the aim to synthesize completely novel data. While this approach reduces memorization, the quality of the samples synthesized by the second model trained on refined synthetic data can be compromised. 
\textcolor{black}{It could be argued that memorization could be mitigated by training generative models on large datasets in federated learning setups \cite{tolle2024federated, kaissis2020secure}. However, depending on the size of the training samples at different imaging centers some data samples can be more prone to memorization.}
Other potential approaches to mitigate memorization can be using differential private diffusion models \cite{DPDMs, Malte2022privacy} or optimizing model capacity \cite{dutt2024memcontrol}.\\
\textcolor{black}{Some studies have also reported memorization in task-specific downstream models \cite{haim2022reconstructing,He2020,sablayrolles2019dj,hartley2023neural}. More importantly, rare cases can be more prone to memorization \cite{hartley2023neural}, which can make patient data identifiable in a much easier manner. Therefore, extra precautions should be taken while sharing trained models where membership inference attacks can be performed or training images can be reconstructed. It is difficult to draw a direct relation between memorization in generative models and downstream task-specific models because in the former memorization would correspond to assigning very high likelihoods to training samples and in the latter memorization would correspond to remembering features of the training data that could be utilized to infer training samples. Nonetheless, both cases are crucial and require more attention. } \\
Patient data memorization in diffusion models can have broad implications in applications of generative AI in medicine. 
In open-data sharing, patients might not be comfortable making their data publicly available, which is one of the core reasons why generative models are deployed for open-data sharing in the first place. Incidentally sharing patient data copies defeats the whole purpose. 
Furthermore, patient data copies among synthetic images can also be potentially traced back to the original patient leading to patient re-identification. 
\textcolor{black}{This can happen in several ways. First data samples accidentally containing patient data on the image can be present in the training data and a recent study on image classification showed that features that are anomalies can be memorized to a greater extent in neural networks \cite{hartley2023neural}. Second, patient data identification can happen using demographic information in public datasets. The risk of identification can be higher in the case of longitudinal data, the models can learn patterns specific to patients and therefore might be more easily identifiable in such cases. Third, identification can also take place in case partial information about a patient is available.}
Packhäuser et al. \cite{packhauser2022reidentification} were able to identify two X-ray images from the same patient acquired at different times even when the patient's conditions altered. Using such approaches, an attacker can use partially available patient information to recover patient data copies among presumed novel synthetic data and recover sensitive clinical information.
Another prominent application of generative models is data expansion diversification \cite{datadiversification}. In data expansion and diversification, generative models are trained to synthesize novel data and complement the training data with synthetic data for data-hungry AI models. We observed that a high percentage of synthetic data were patient data copies, especially in the 3D datasets. This also brings the data diversification and expansion application of generative models into question.
\section{Methods}
\subsection{Latent Diffusion Models}
\begin{figure}
\centering
\includegraphics[width=0.8\linewidth]{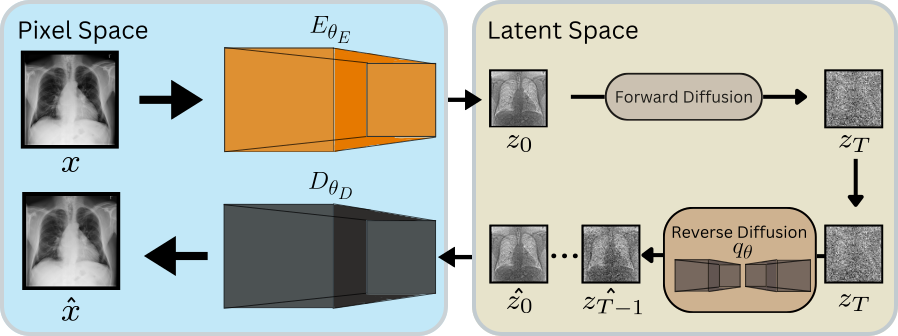}
\caption{Latent Diffusion Models project data onto a lower dimensional latent space and perform gradual denoising followed by projection back onto the pixel space.} 
\label{fig:LDM}
\end{figure}

Latent diffusion models (LDMs) belong to a family of likelihood-based generative models that are designed to learn data distribution $p(x)$ through a gradual denoising process in a low dimensional latent space (see Fig. \ref{fig:LDM}) \cite{LDM}. 
The latent space is learned through an autoencoder. Given an image $x$, the encoder $E_{\theta_E}$ projects $x$ onto its low dimensional latent representation $z$, followed by a back projection onto the original pixel space as $\hat{x}$. 
This latent space project reduces computational complexity and enables application on high-resolution images \cite{LDM}.
The autoencoder is trained using a reconstruction loss ($\mathcal{L}_{rec}$) that enforces the model to learn a meaningful compressed representation, and adversarial ($\mathcal{L}_{adv}$) and perceptual ($\mathcal{L}_{prec}$) losses for enhanced perceptual quality of the reconstructed image. The cumulative loss ($\mathcal{L}_{com}$) can be expressed as a weighted summation of the reconstruction loss with weighting $\lambda_{rec}$, adversarial loss with weighting $\lambda_{adv}$, and perceptual loss with weighting $\lambda_{perc}$\\
After learning the latent space, deep diffusion models undergo training in the latent space. 
Deep diffusion models are trained with the aim to minimize the upper variational bound of the negative log-likelihood of the data distribution  $-\log (p(z))$ \cite{DDPM}. 
Training a diffusion model consists of two steps. The first step constitutes a forward diffusion process where normally distributed noise is added to the latent representation $z$ of images $x$. This process is performed in small increments ($\delta t$) with a variance schedule of ($\beta_{t}$), resulting in noisy representations $z_{t}$ at every value of $t$. At any time $t$, $q\left( z_{t}|z_{t-1} \right)$ is modelled as a normal distribution with mean $\sqrt{1-\beta_{t}}z_{t-1}$ and variance $\beta_{t}$. 
The second step consists of a reverse diffusion process aimed at learning $q\left( z_{t-1}|z_{t} \right)$. Unlike $q\left( z_{t}|z_{t-1} \right)$, $q\left( z_{t-1}|z_{t} \right)$ does not have a closed form expression and is typically estimated using a deep neural network $\hat{q_{\theta}}\left( z_{t-1}|z_{t} \right)$ at various different values of $t$.\\
Once the models are trained, they can be used to generate samples by initiating from random noise $z_{T}\sim \mathcal{N}\left( 0,\mathrm{I} \right)$ and performing sequential denoising to obtain a new sample $z^{'}_0$ in the latent space. This new latent sample can be projected back to the pixel space using the decoder $D_{\theta_D}$.
\subsection{Memorization Assessment} \label{methods:mem_assesment}
Despite the ability of latent diffusion models to generate high-quality and realistic samples, the affinity of such models to memorize patient data and synthesize it has received little attention \cite{dar2023investigating, dar2024epoch, akbar2023beware}. 
Akbar et al. \cite{akbar2023beware} defined memorization as a phenomenon where generative models synthesize patient data copies and defined copies as synthesized samples that are identical to training samples. Dar et al. \cite{dar2023investigating} further expanded the definition of copies to further include variations such as rotation, flipping, and minor changes in contrast. Here we adhere to this expanded definition, and drawing inspiration from Fernandez et al. \cite{fernandez2023privacy} we formally define memorization as follows:\\ 
A training data sample $x$ is considered to be $(l,\rho)-memorized$ by a generative model $G_{\theta}$ if $l(x, \upsilon (\hat{x})) \geq \tau $, where $\hat{x}$ is a sample extracted from $G_{\theta}$ using sampling algorithm $A$, $\upsilon$ corresponds to minor variations such rotation, flipping and slight changes in contrast, $l$ is the similarity between the samples, and $\tau$ is a threshold level. Under such conditions, $\hat{x}$ is defined as a copy of $x$.\\
\subsubsection{Contrastive Learning}
\begin{figure}
\centering
\includegraphics[width=0.8\linewidth]{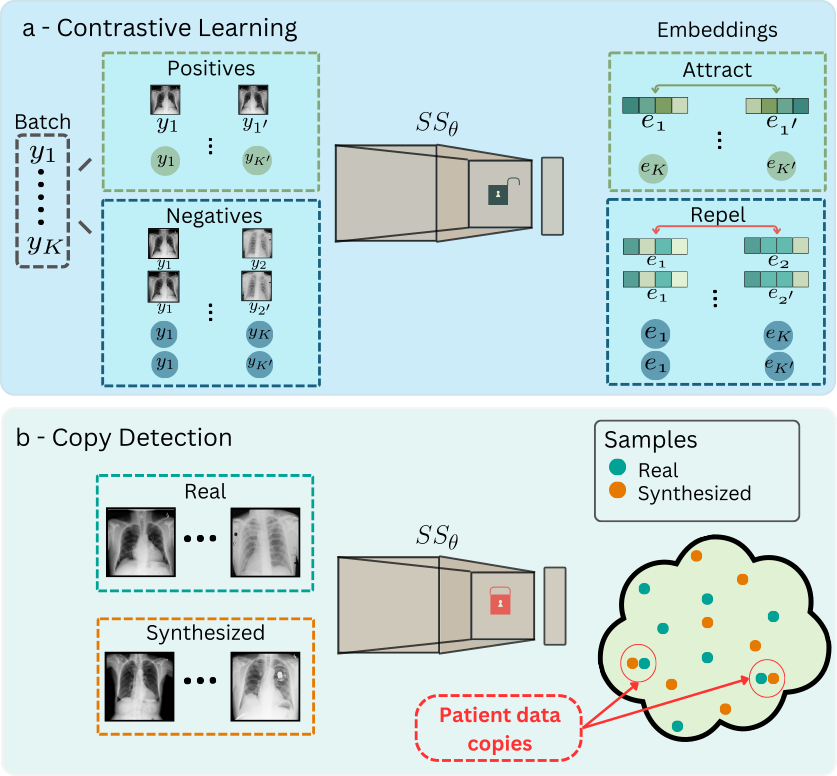}
\caption{a - Self-supervised model is trained to project images onto a lower dimensional embedding space where each sample within a batch is brought closer to its variation and pushed far away from all other samples in the batch. b - Patient data copies are detected by projecting all training and synthetic samples onto the embedding space and identifying the training synthetic pairs lying close to each other.} 
\label{fig:mem-assessment}
\end{figure}

One naive way to detect patient data copies is to compare each synthesized sample with all training samples and select the samples showing a similarity level greater than the threshold $\tau$ as copies. However, this approach is computationally inefficient and is not suitable for detecting copies that are variations of patient images. Accordingly, we utilized self-supervised models ($SS_{\theta}$) that project images into a lower dimensional embedding space and used a contrastive learning approach \cite{simclr} to bring each training sample closer to its variation and push away from other samples (Fig. \ref{fig:mem-assessment}). 
The rationale behind this approach is that copies would lie closer to the training samples and novel samples would be far away. The details of the model are mentioned in Section \ref{suppmethods:self_sup} \\
\subsubsection{Copy Detection} \label{methods:copy_detection}
For the detection of patient data copies, $SS_{\theta}$ was first utilized to obtain embeddings of all training, validation, and synthetic samples. 
Next, Pearson's correlation coefficient was computed between all pairs of training-validation embeddings ($\rho_{tr-val}$), and training-synthetic embeddings ($\rho_{tr-syn}$).
Afterwards, for each training embedding, the closest validation embedding was selected to form a distribution of their correlation values ($\rho_{NN-val}$). A threshold value ($\tau$) was then defined as $95th$ percentile of $\rho_{NN-val}$.
Finally, for each training embedding the closest synthetic embedding was selected ($\rho_{NN-syn}$), and training samples with $\rho_{NN-syn}$ greater than $\tau$ were categorized as memorized. 
Algorithm \ref{suppalg:copy_detection} demonstrates the copy detection process via a pseudo code.
\subsection{ Training and Evaluation Procedures}
\begin{figure}[h]
\includegraphics[width=1\linewidth]{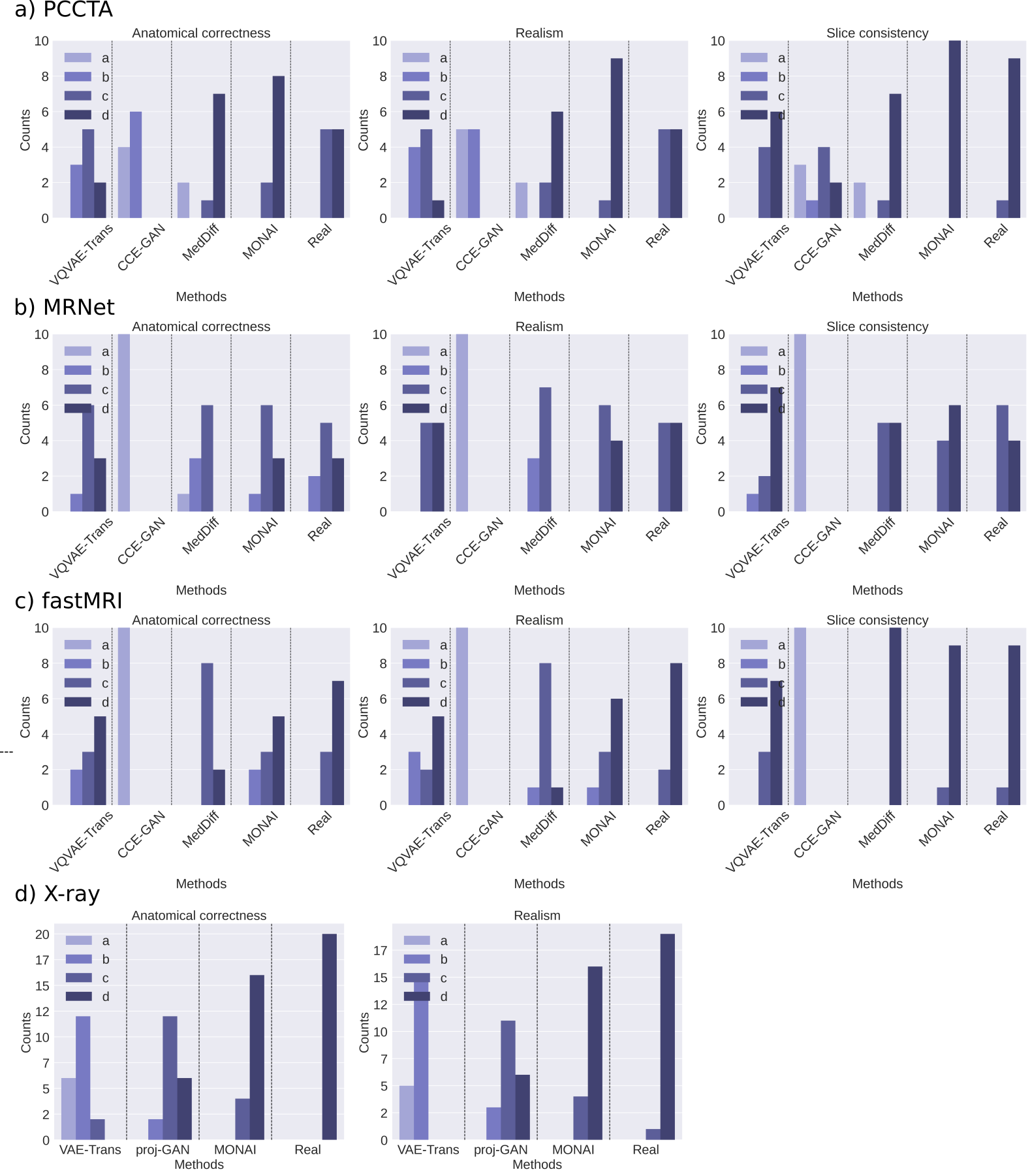}
\caption{Expert evaluation scores on a) PCCTA, b) MRNet, c) fastMRI, and d) X-ray datasets (for criteria please see section 4.4). MONAI consistently outperforms other models across all datasets and criteria, except for Realism in the MRNet dataset, where VQVAE-Trans shows a slight edge, and Anatomical Correctness in (MRNet, fastMRI), where VQVAE-Trans has similar performance levels. Most images synthesized by MedDiff are generally satisfactory, with minor inaccuracies in terms of realism and anatomical correctness. While VQVAE-Trans performs reasonably well on 3D datasets, it struggles in the 2D X-ray dataset. In contrast, GANs excel on the 2D X-ray dataset but fail on the 3D datasets. Many of the Real samples in MRNet are also sometimes categorized into class c. This could be attributed to the severe artifacts present in the MRNet dataset.} 
\label{suppfig:rad1}
\end{figure}

\begin{figure}[h]
\includegraphics[width=1\linewidth]{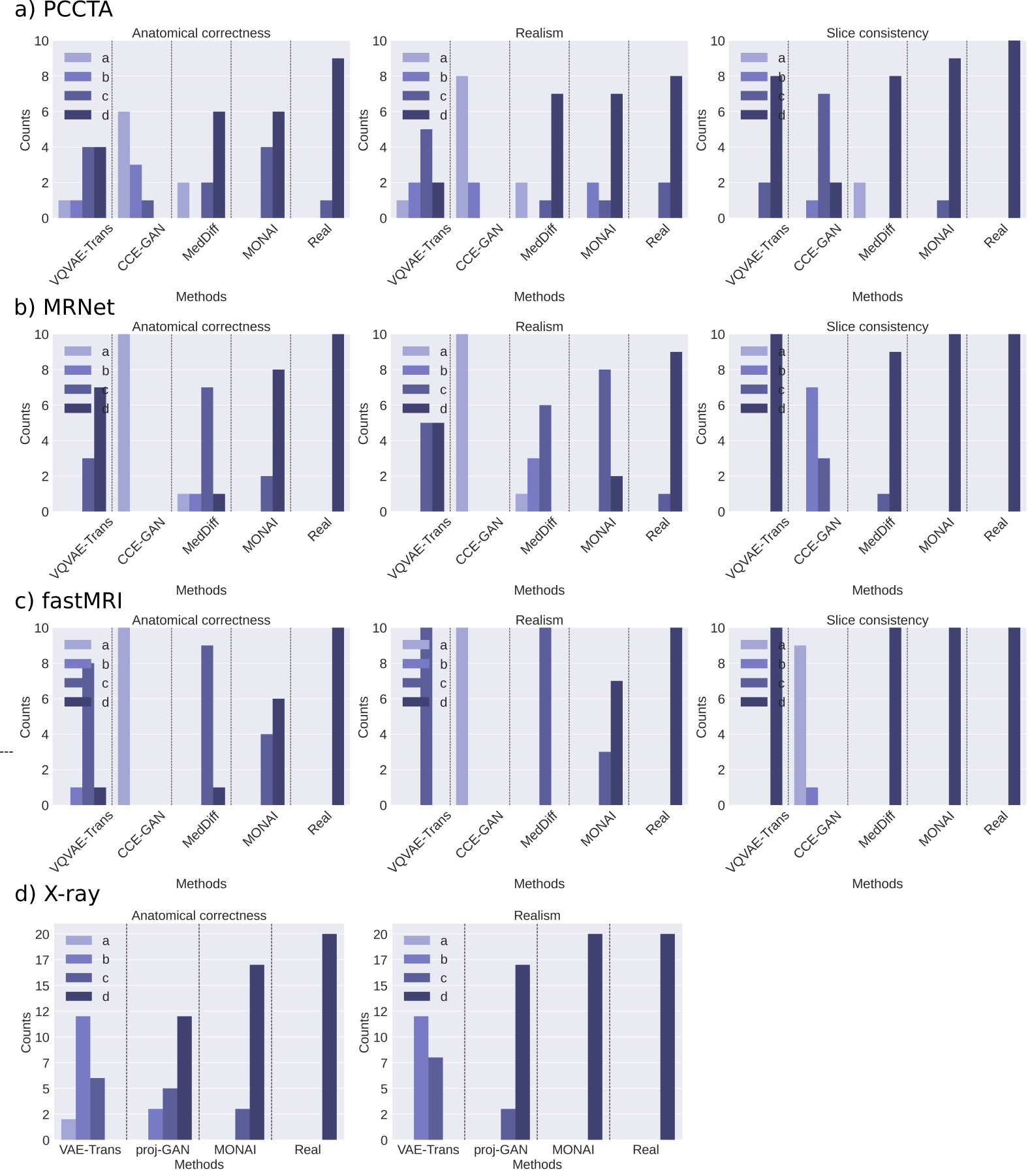}
\caption{Expert evaluation scores on a) PCCTA, b)MRNet, c) fastMRI, and d)X-ray datasets (for criteria please see section 4.4). MONAI consistently outperforms other models across all datasets and criteria, except for Realism in the MRNet dataset, where VQVAE-Trans shows a slight edge. Most images synthesized by MedDiff are generally satisfactory, with minor inaccuracies in terms of realism and anatomical correctness. While VQVAE-Trans performs reasonably well on 3D datasets, it struggles with the 2D X-ray dataset. In contrast, GANs excel on the 2D X-ray dataset but fail on the 3D datasets.} 
\label{suppfig:rad2}
\end{figure}

\paragraph{Latent Diffusion Models:}
For the 3D datasets, two models were considered (MedDiff \cite{Khader2023} and MONAI\cite{pinaya2023monai}). We opted for these models because they are some of the most widely used repositories for medical image synthesis using latent diffusion models. Training procedures, network architectures, and hyperparameters for MedDiff were adopted from Khader et al. \cite{Khader2023} (\url{https://github.com/FirasGit/medicaldiffusion}) and for MONAI were adopted from the online repository of Pinaya et al. \cite{pinaya2023monai} (\url{https://github.com/Project-MONAI/tutorials/blob/main/generative/3d_ldm/config/config_train_32g.json}). The only exception was the batch size of the models, which was modified to 8 for the autoencoder and 20 for the diffusion model. MedDiff consisted of a vector quantized generative adversarial network (VQ-GAN) with 3D convolutions as the autoencoder and MONAI consisted of a variational autoencoder (VAE) with 3D convolutions as the autoencoder. All 3D-LDMs were trained for roughly 150k iterations, and the number of sampling time steps was set to 300 in MedDiff and 1000 in MONAI. 
For the 2D X-ray dataset, training procedures, network architectures, and hyperparameters were adopted from an online repository (\url{https://github.com/Warvito/generative_chestxray}) built on the MONAI framework \cite{pinaya2023monai}, and is referred to as MONAI-2D. MONAI-2D consisted of VAE with 2D convolutions as the autoencoder. All 2D-LDMs were trained for 200k iterations unless specified, and the number of sampling time steps was set to 1000.\\
\textcolor{black}{\paragraph{Non-Diffusion Generative Models:}
In the 3D datasets, CCE-GANs and VQVAVE-transformers (VQVAE-Trans) were utilized. For CCE-GAN, all training and synthesis procedures were adapted from \url{https://github.com/ShiboXing/Improved-3dbraingen/CCE_GAN_3.ipynb}. The only exception was the architectures, where the number of sampling and downsampling layers were modified based on the volume dimensions within each dataset. 
For the training of VQVAE-Trans, all training procedures were adapted from \url{https://github.com/marksgraham/transformer-ood} and synthesis procedures were adapted from \url{https://github.com/Project-MONAI/GenerativeModels/tree/main/tutorials/generative/2d_vqvae_transformer}. The only exceptions were the number of downsampling and upsampling operations, which were set to 3 in the PCCTA dataset.
In the 2D datasets, projected-GAN (proj-GAN) and VQVAE-Transformers were used. For proj-GAN, all training and synthesis procedures were adapted from \url{https://github.com/autonomousvision/projected-gan}. For 2D VQVAE-Trans, all training and synthesis procedures were adapted from \url{https://github.com/Project-MONAI/GenerativeModels/tree/main/tutorials/generative/2d_vqvae_transformer}. The only exceptions were the number of downsampling and upsampling operations, which were set to 3 instead of 2 due due memory limitations and the number of channels for the first downsampling operation was set to 128.}
\paragraph{Self-Supervised Models:}
In 3D-self supervised models, the architecture of the encoder was modified from the encoder of VQ-VAE in MedDiff. The network was used to reduce the input dimensions to $4^3$ and 8 channels, followed by flattening and using dense layers to reduce the dimensions to a $32\times1$ vector, \textcolor{black}{making the embedding size equal to 32}. The training procedures were adopted from Dar et al. \cite{dar2023investigating}
In 2D self-supervised models, the model was adopted from Packhäuser et al. \cite{packhauser2022reidentification} with minor modifications in architecture. The last classification layer was replaced by a dense layer having dimensions of $128\times1$ as output, \textcolor{black}{making the embedding size equal to 128}. The training procedures were adopted from Dar et al. \cite{dar2024epoch}. \\
\paragraph{Evaluation Metrics:}
In all synthesized datasets FID and MS-SSIM were adopted from the MONAI repository (\url{https://github.com/Project-MONAI/GenerativeModels/tree/main/generative/metrics}). In 3D datasets, FID was calculated between features extracted from whole datasets. The features for FID calculation were extracted using a pre-trained model adapted from Chen et al \cite{fid-3d}. In the 2D dataset, FID was calculated in batches with a batch size of 256 and averaged afterwards. The features for FID calculation were extracted using a pre-trained model adopted from Cohen et al. \cite{fid-2d}. The reported average MS-SSIM values were calculated by computing MS-SSIM values between each synthetic sample and a randomly selected synthetic sample and averaging them. 
\textcolor{black}{\subsection{Radiological evaluation}\label{methods:rad_eval_crit}
Images in section \ref{results:mem_gans_ae} were evaluated by expert radiologists based on the following criteria:\\
\textbf{Realism}: a) Overall not recognizable as CT/MRI/Xray b) Overall unrealistic but generally recognizable as CT/MRI/Xray c) Overall realistic and only minor unrealistic areas d) Can't tell whether fake or not.\\
\textbf{Anatomical correctness}: a) Anatomic region not recognizable b) Anatomic region recognizable but major parts of the images exhibit anatomic incorrectness c) Only minor anatomic incorrectness d) Anatomic features are correct.\\
\textbf{Consistency between slices} (only applies to 3D volumes): a) No consistent slices b) Only a few slices are consistent c) Majority of the slices are consistent d) All slices are consistent.\\
The scoring task was distributed among three expert radiologists. Expert 1 scored images in PCCTA, MRNet, and fastMRI datasets (Supp. Fig. \ref{suppfig:rad1}), expert 2 scored images in the X-ray dataset (Supp. Fig. \ref{suppfig:rad1}), and expert 3 scored images in all four datasets (Supp. Fig. \ref{suppfig:rad2}). 
}
\subsection{Datasets}
PCCTA data were acquired at the University Hospital Mannheim on a Siemens Naeotom Alpha scanner. Ethics approval was granted by the Ethics Committee of Ethikkommission II at Heidelberg University (ID 2021-659). Data were acquired from 64 patients, with coronary artery plaques detected and annotated by an expert radiologist. Sub-volumes of sizes 64\textsuperscript{3} surrounding coronary artery plaques were cropped, resulting in 300 sub-volumes. Among these, 242 sub-volumes from 51 subjects were reserved for training, and 58 sub-volumes from 13 subjects were reserved for validation.\\
In the MRNet dataset, 1130 sagittal scans from 1088 patients were used in this study, where 904 volumes were reserved for training and 226 for validation. All volumes were cropped or zero-padded to have sizes of 256\textsuperscript{2}x32. The data loader and train-validation split for the MRNet dataset was directly adopted from \url{https://github.com/FirasGit/medicaldiffusion/blob/master/dataset/mrnet.py}.\\
\textcolor{black}{In the fastMRI dataset, Dicom images of brain MRI scans were obtained from the NYU fastMRI dataset \cite{fastmri1,fastmri2}. Afterwards, all images were resampled to have sizes of 256\textsuperscript{2}x16. The first 1k samples from the first 443 subjects (as shared publicly) were reserved for training. The last 10k images from 4381 patients (as shared publicly) were initially reserved for validation. However, to reduce computational complexity the first 1k images (443 subjects) of the selected samples were only utilized for validation.} \\
In the X-ray dataset, 10k samples from the first 2587 patients (as shared publicly) were reserved for training and 10k samples from the last 3609 patients (as shared publicly) were reserved for validation. To reduce computational complexity, all images were sampled to sizes of 512\textsuperscript{2}.
\paragraph{Data Availability:}
This study utilizes \textcolor{black}{four} datasets. The MRNet, \textcolor{black}{fastMRI, and} X-ray datasets are publicly available at \url{https://stanfordmlgroup.github.io/competitions/mrnet/}, \url{https://fastmri.med.nyu.edu/}, and \url{https://www.kaggle.com/datasets/nih-chest-xrays/data} respectively. The synthesized MRNet, \textcolor{black}{fastMRI}, and X-ray images will be shared upon request and proof that the requester has access to the corresponding public datasets. \\
The in-house dataset, cannot be made publicly available due to restrictions imposed by the University Hospital Mannheim, where the data were acquired.
\paragraph{Code Availability:}
In adherence to the FAIR principles (findability, accessibility, interoperability, and reusability) in scientific research, all code utilized in this work is publicly available.\\
The LDMs were trained using code from public repositories. Specifically, MedDiff was trained using code from \url{https://github.com/FirasGit/medicaldiffusion}. MONAI was trained using code from \url{https://github.com/Project-MONAI/}, and MONAI-2D utilized code from \url{https://github.com/Warvito/generative_chestxray}. \\
Our code for data pre-processing and training and evaluation of self-supervised models for copy detection is made publicly available at \url{https://github.com/Cardio-AI/memorization-ldm}.
\paragraph{Acknowledgement:}
S.U.H.D. is supported through state funds approved by the State Parliament of Baden-Württemberg for the Innovation Campus Health + Life Science Alliance Heidelberg Mannheim.\\
S.F. is supported by the German Federal Ministry of Education and Research (SWAG, 01KD2215C), the German Cancer Aid (DECADE, 70115166) and the German Research Foundation (504101714).\\
J.N.K. is supported by the German Cancer Aid (DECADE, 70115166), the German Federal Ministry of Education and Research (PEARL, 01KD2104C; CAMINO, 01EO2101; SWAG, 01KD2215A; TRANSFORM LIVER, 031L0312A; TANGERINE, 01KT2302 through ERA-NET Transcan), the German Academic Exchange Service (SECAI, 57616814), the German Federal Joint Committee (TransplantKI, 01VSF21048) the European Union’s Horizon Europe and innovation programme (ODELIA, 101057091; GENIAL, 101096312), the European Research Council (ERC; NADIR, 101114631) and the National Institute for Health and Care Research (NIHR, NIHR203331) Leeds Biomedical Research Centre. The views expressed are those of the author(s) and not necessarily those of the NHS, the NIHR or the Department of Health and Social Care. This work was funded by the European Union. Views and opinions expressed are however those of the author(s) only and do not necessarily reflect those of the European Union. Neither the European Union nor the granting authority can be held responsible for them. \\
S.E. is supported by BMBF-SWAG Project 01KD2215D, Carl-Zeiss-Stiftung within the Multi-dimensionAI consortium, and Informatics for Life project through the Klaus Tschira Foundation. \\
The authors also gratefully acknowledge the data storage service SDS@hd supported by the Ministry of Science, Research and the Arts  Baden-Württemberg (MWK) and the German Research Foundation (DFG) through grant INST 35/1314-1 FUGG and INST 35/1503-1 FUGG. The authors also acknowledge support by the state of Baden-Württemberg through bwHPC
and the German Research Foundation (DFG) through grant INST 35/1597-1 FUGG.
\paragraph{Conflict of interest:}
S.O.S. is the director of the Dept. of Radiology and Nuclear Medicine at the University Medical Center in Mannheim, which has research agreements with Siemens Healthineers.\\
B.B. is the founder and CEO of LernRad GmbH and a speaker at Bureau Bayer Vital GmbH.\\
S.F. has received honoraria by MSD and BMS.\\
D.T. holds shares in StratifAI GmbH and has received honoraria for lectures from Bayer.\\
J.N.K. declares consulting services for Owkin, France; DoMore Diagnostics, Norway; Panakeia, UK; AstraZeneca, UK; Scailyte, Switzerland; Mindpeak, Germany; and MultiplexDx, Slovakia. Furthermore he holds shares in StratifAI GmbH, Germany, has received a research grant by GSK, and has received honoraria by AstraZeneca, Bayer, Eisai, Janssen, MSD, BMS, Roche, Pfizer and Fresenius.\\
S.E. has received honoraria by Boehringer Ingelheim.
\paragraph{Inclusion and Ethics:}
This research was performed via a collaboration between researchers from several institutes, and all authors fulfilled the authorship criterion. The contributions of the authors are mentioned as follows:\\
S.U.H.D. and S.E. developed the ideas for the study and initiated the project.\\
J.K., I.A., and S.O.S. acquired the photon-counting computed tomography angiography dataset, performed pre-processing, and labeled the regions containing plaques.\\
\textcolor{black}{I.A., R.M.S., and F.C.L. performed expert radiological evaluation on the images.}\\
S.U.H.D., M.S., and S.E. designed methodology and experiments.\\
S.U.H.D. and M.S. performed all experiments, and wrote the codes for all experiments.\\
S.U.H.D., M.S., J.N.K, and S.E. interpreted the results.\\
S.U.H.D., \textcolor{black}{M.S.}, and S.E. wrote the manuscript.\\
T.P., S.O.S, N.F., B.B., S.F., D.T., J.N.K, and S.E. guided the study and provided feedback throughout the manuscript preparation.
\bibliographystyle{IEEEtran} 
\bibliography{references} 
\newpage
\setcounter{page}{1}
\beginsupplement
\section*{\makebox[\textwidth][c]{Supplementary Material}}
\section{Supplementary Methods}
\subsection{Self-Supervised Model Training}\label{suppmethods:self_sup}
Consider a batch $B=[y_{1}, y_{2},..., y_{K}]$ containing $K$ samples, where $y_i$ corresponds to $i^{th}$ sample. After obtaining variation $y_{i}'$ for each sample $y_i$, the modified batch can then be represented as  $B'=[y_{1}, y_{1}', y_{2}, y_{2}',..., y_{K}, y_{K}']$. 
One straightforward way is to form a positive $(y_i,y_i')$ and a negative pair $(y_i,y_j)$ for each sample $y_i$.
However, we observed that such an approach was unable to efficiently push samples within a negative pair away from each other. 
Therefore we increased the number of negative pairs for each sample, such that for each sample  $y_i$ negative pairs were formed using all other samples in the batch, making $2(K-1)$ negative pairs per sample.\\
The self-supervised model $SS_{\theta}$ model was trained using the normalized temperature-scaled cross entropy (NT-Xent) loss \cite{simclr}. First, $SS_{\theta}$ was used to obtain embeddings $E=[e_{1}, e_{1}', e_{2}, e_{2}',..., e_{K}, e_{K}']$ of all samples within $B'$. For each $ith$ sample it was then updated based on the NT-Xent loss function, expressed as follows:
\begin{equation}\label{suppeqn:ntxent}
\begin{aligned}
\mathcal{L}_{i} = -\log \frac{ e^{(s_{e_{i}, e_{i}'}/\tau)}}
{\sum_{j=1}^{2K}\mathds{1}_{[j \neq i]}e^{(s_{e_{i}, e_{j}}/\tau)}} +  \frac{ e^{(s_{e_{j}', e_{j}}/\tau)}}  {\sum_{j=1}^{2K}\mathds{1}_{[j \neq i']} e^{(s_{e_{i'}, e_{j}}/\tau)}} 
\end{aligned}
\end{equation}
Here $s_{e_{i'}, e_{j}}$ is cosine similarity $ith$ and $jth$ embeddings and $\mathds{1}_{[j \neq i]}$ is the indicator function which is 1 when $j \neq i$ and 0 when $j = i$.
\newpage
\begin{algorithm}
\DontPrintSemicolon
\KwInput{$E_{tr}= [e_{tr}^{1},...,e_{tr}^{N_{tr}} ]^T \in \mathbb{R}^{N_{tr} \times L}: N_{tr}$ training embeddings of length $L$\\
$E_{val}= [e_{val}^{1},...,e_{val}^{N_{val}}]^T\in \mathbb{R}^{N_{val} \times L}: N_{val}$ validation embeddings\\
$E_{syn}= [e_{syn}^{1},...,e_{syn}^{N_{syn}}]^T\in \mathbb{R}^{N_{syn} \times L}: N_{syn}$ synthetic embeddings\\
$corr(. , .)$: Pearson's correlation between inputs\\
$percentile(. , u)$: $uth$ percentile of input vector \\
$ind()$: Indices of True values
}
\KwOutput{$ID_{cop}$ : Indices of memorized samples}
$\rho_{tr-val} = corr(E_{tr}, E_{val})\in \mathbb{R}^{N_{tr} \times N_{val}}$ \tcp*{Pairwise correlations between embeddings} 
$\rho_{tr-syn} = corr(E_{tr}, E_{val})\in \mathbb{R}^{N_{tr} \times N_{syn}}$  \\
$\rho_{NN-val} = max (\rho_{tr-val} )\in \mathbb{R}^{N_{tr}}$ \tcp*{Nearest neighbor selection for each training embedding} 
$\rho_{NN-syn} = max (\rho_{syn-val} )\in \mathbb{R}^{N_{tr}}$ \\
$\tau = percentile(\rho_{NN-val}, 95)$ \tcp*{Threshold based on 95 percentile}
$ID_{cop} = ind(\rho_{NN-syn} \geq \tau)$ \tcp*{Indices of training samples that are memorized}
\caption{Copy detection}
\label{suppalg:copy_detection}
\end{algorithm}
\newpage
\section{Supplementary Figures}
\newpage
\newpage
\newpage
\newpage
\newpage
\newpage
\newpage

\newpage
\newpage
\newpage
\newpage
\newpage
\newpage
\newpage
\newpage

\clearpage

\newpage
\newpage
\newpage
\clearpage
\section{Supplementary Tables}
\begin{table}[h]
    \centering
    \caption{Percentage of training data memorized and synthesized samples that are copies in models trained on different datasets.}
    \begin{tabular}{@{}lcccccc@{}}
        \toprule
        \textbf{Dataset (N$_{\text{train}}$)} & \textbf{Method} & \textbf{N$_{\text{mem}}$ (\%)} & \textbf{N$_{\text{copy}}$ (\%)} \\
        \midrule
        \multirow{4}{*}{PCCTA (242)} 
        & MedDiff & 43.8 & 91.7 \\
        & Monai & 40.5 & 83.1 \\
        & CC-GAN & 5.4 & 5.0 \\
        & VAE Trans & 49.6 & 66.1 \\
        \midrule
        \multirow{4}{*}{MRNet (904)} 
        & MedDiff & 40.2 & 76.1 \\
        & Monai & 48.3 & 87.4 \\
        & CC-GAN & 12.3 & 13.3 \\
        & VAE Trans & 58.3 & 83.3 \\
        \midrule
        \multirow{4}{*}{fastMRI (1k)} 
        & MedDiff & 24.8 & 37.3 \\
        & Monai & 30.8 & 51.0 \\
        & CC-GAN & 1.3 & 1.7 \\
        & VAE Trans & 40.2 & 57.5 \\
        \midrule
        \multirow{3}{*}{Xray (10k)} 
        & Monai & 32.6 & 54.5 \\
        & pGAN & 3.2 & 4.5 \\
        & VAE Trans & 2.6 & 3.5 \\
        \bottomrule
    \end{tabular}
    \label{supptab:comparison_methods}
\end{table}
\newpage
\begin{table}[h]
    \centering
    \caption{Comparison of percentage $N_{\text{mem}}$ and $N_{\text{copy}}$ across different batch sizes used for training of self-supervised methods for copy detection.}
    \begin{tabular}{@{}lcccccc@{}}
        \toprule
        \textbf{Dataset (N$_{\text{train}}$)} & \textbf{Batch Size} & \textbf{Method} & \textbf{N$_{\text{mem}}$} (\%) & & \textbf{N$_{\text{copy}}$} (\%) \\
        \midrule
        \multirow{8}{*}{PCCTA (242)} 
        & 2 
            & MedDiff & 47.9 & & 87.6 \\
        & & Monai & 40.1 & & 74.0 \\
        \cmidrule{2-6}
        & 5 
            & MedDiff & 44.2 & & 90.5 \\
        & & Monai & 37.6 & & 76.0 \\
        \cmidrule{2-6}
        & 10 
            & MedDiff & 43.8 & & 91.7 \\
        & & Monai & 40.5 & & 83.1 \\
        \cmidrule{2-6}
        & 20 
            & MedDiff & 42.6 & & 91.7 \\
        & & Monai & 45.0 & & 86.0 \\
        \midrule
        \multirow{8}{*}{MRNet (904)} 
        & 2 
            & MedDiff & 29.9 & & 50.7 \\
        & & Monai & 41.4 & & 70.1 \\
        \cmidrule{2-6}
        & 5 
            & MedDiff & 38.5 & & 69.4 \\
        & & Monai & 47.4 & & 81.6 \\
        \cmidrule{2-6}
        & 10 
            & MedDiff & 40.2 & & 76.1 \\
        & & Monai & 48.3 & & 87.4 \\
        \cmidrule{2-6}
        & 20 
            & MedDiff & 40.5 & & 78.0 \\
        & & Monai & 48.6 & & 89.8 \\
        \midrule
        \multirow{8}{*}{fastMRI (1k)} 
        & \multirow{2}{*}{2} 
            & MedDiff & 7.2 & & 9.8 \\
        & & Monai & 15.4 & & 22.3 \\
        \cmidrule{2-6}
        & \multirow{2}{*}{5} 
            & MedDiff & 24.6 & & 34.4 \\
        & & Monai & 27.7 & & 42.8 \\
        \cmidrule{2-6}
        & \multirow{2}{*}{10} 
            & MedDiff & 24.8 & & 37.3 \\
        & & Monai & 30.8 & & 51.0 \\
        \cmidrule{2-6}
        & \multirow{2}{*}{20} 
            & MedDiff & 21.3 & & 30.5 \\
        & & Monai & 29.0 & & 45.8 \\
        \midrule
        \multirow{3}{*}{Xray (10k)} 
        & 32  & \multirow{3}{*}{Monai} & 10.8 & & 13.3 \\
        & 64  & & 32.6 & & 54.5 \\
        & 128 & & 32.6 & & 55.3 \\
        \bottomrule
    \end{tabular}
    \label{supptab:comparison_batch_size}
\end{table}
\newpage
\begin{table}[h]
    \centering
    \caption{Comparison of percentage $N_{\text{mem}}$ and $N_{\text{copy}}$ across different embedding sizes in self-supervised methods for copy detection.}
    \begin{tabular}{@{}lcccccc@{}}
        \toprule
        \textbf{Dataset (N$_{\text{train}}$)} & \textbf{Embedding Size} & \textbf{Method} & \textbf{N$_{\text{mem}}$} (\%) & & \textbf{N$_{\text{copy}}$} (\%) \\
        \midrule
        \multirow{6}{*}{PCCTA (242)} 
        & \multirow{2}{*}{16} 
            & MedDiff & 45.5 & & 90.5 \\
        & & Monai & 38.0 & & 72.3 \\
        \cmidrule{2-6}
        & \multirow{2}{*}{32} 
            & MedDiff & 43.8 & & 91.7 \\
        & & Monai & 40.5 & & 83.1 \\
        \cmidrule{2-6}
        & \multirow{2}{*}{64} 
            & MedDiff & 44.2 & & 90.9 \\
        & & Monai & 45.0 & & 81.4 \\
        \midrule
        \multirow{6}{*}{MRNet (904)} 
        & \multirow{2}{*}{16} 
            & MedDiff & 37.3 & & 67.6 \\
        & & Monai & 46.0 & & 80.2 \\
        \cmidrule{2-6}
        & \multirow{2}{*}{32} 
            & MedDiff & 40.2 & & 76.1 \\
        & & Monai & 48.3 & & 87.4 \\
        \cmidrule{2-6}
        & \multirow{2}{*}{64} 
            & MedDiff & 40.2 & & 77.7 \\
        & & Monai & 48.7 & & 87.2 \\
        \midrule
        \multirow{8}{*}{fastMRI (1k)} 
        & \multirow{2}{*}{16} 
            & MedDiff & 25.5 & & 37.6 \\
        & & Monai & 30.0 & & 51.2 \\
        \cmidrule{2-6}
        & \multirow{2}{*}{32} 
            & MedDiff & 24.8 & & 37.3 \\
        & & Monai & 30.8 & & 51.0 \\
        \cmidrule{2-6}
        & \multirow{2}{*}{64} 
            & MedDiff & 24.3 & & 35.4 \\
        & & Monai & 28.0 & & 44.3 \\
        \midrule
        \multirow{4}{*}{Xray (10k)} 
        & 32  & \multirow{4}{*}{Monai} & 29.7 & & 50.5 \\
        & 64  & & 29.4 & & 49.0 \\
        & 128 & & 32.6 & & 54.5 \\
        & 256 & & 31.4 & & 51.0 \\
        \bottomrule
    \end{tabular}
    \label{supptab:comparison_embedding_size}
\end{table}
\newpage
\begin{table}[h]
    \centering
    \caption{Comparison of percentage $N_{\text{mem}}$ and $N_{\text{copy}}$ across different numbers of synthetic samples ($N_{\text{synth}}$).}
    \begin{tabular}{@{}lccccc@{}}
        \toprule
        \textbf{Dataset (N$_{\text{train}}$)} & \textbf{N$_{\text{synth}}$} & \textbf{N$_{\text{mem}}$ (\%)} & & \textbf{N$_{\text{copy}}$ (\%)} \\
        \midrule
        \multirow{3}{*}{fastMRI (1k)} 
        & 0.5k & 20.8 & & 48.8 \\
        & 1k   & 31.5 & & 49.4 \\
        & 2k   & 49.7 & & 49.5 \\
        \midrule
        \multirow{3}{*}{Xray (10k)} 
        & 5k   & 21.0 & & 54.0 \\
        & 10k  & 32.7 & & 54.7 \\
        & 20k  & 46.0 & & 54.3 \\
        \bottomrule
    \end{tabular}
    \label{supptab:comparison_synthetic_samples_rates}
\end{table}
\newpage
\begin{table}[h]
    \centering
    \caption{Comparison of percentage $N_{\text{mem}}$ and $N_{\text{copy}}$ across different numbers of synthetic samples ($N_{\text{synth}}$) when the target data is not the training dataset, but a dummy dataset. Ideally, all entries should be zero.}
    \begin{tabular}{@{}lccccc@{}}
        \toprule
        \textbf{Dataset (N$_{\text{train}}$)} & \textbf{N$_{\text{synth}}$} & \textbf{N$_{\text{mem}}$ (\%)} & & \textbf{N$_{\text{copy}}$ (\%)} \\
        \midrule
        \multirow{3}{*}{fastMRI (1k)} 
        & 0.5k & 1.2 & & 1.4 \\
        & 1k   & 2.2 & & 2.0 \\
        & 2k   & 4.6 & & 5.0 \\
        \midrule
        \multirow{3}{*}{Xray (10k)} 
        & 5k   & 3.8 & & 7.9 \\
        & 10k  & 5.9 & & 7.9 \\
        & 20k  & 9.5 & & 8.1 \\
        \bottomrule
    \end{tabular}
    \label{supptab:comparison_synthetic_samples_rates_dummy}
\end{table}
\newpage
\begin{table}[h]
    \centering
    \caption{Comparison of percentage $N_{\text{mem}}$ and $N_{\text{copy}}$ across different numbers of training samples ($N_{\text{train}}$) when no. of training epochs are fixed.}
    \begin{tabular}{@{}lccccc@{}}
        \toprule
        \textbf{Dataset }(\textbf{N$_{\text{epoch}}$}) & \textbf{N$_{\text{train}}$} & \textbf{N$_{\text{mem}}$ (\%)} & & \textbf{N$_{\text{copy}}$ (\%)} \\
        \midrule
        \multirow{3}{*}{fastMRI (2k)} 
        & 1k & 14.3 & & 24.8 \\
        & 2k   & 12.8 & & 29.8 \\
        & 5k   & 7.4 & & 32.4 \\
        \midrule
        \multirow{3}{*}{Xray (3k)} 
        & 5k   & 54.7 & & 68.6 \\
        & 10k  & 33.2 & & 51.4 \\
        & 20k  & 17.7 & & 39.9 \\
        \bottomrule
    \end{tabular}
    \label{supptab:comparison_Ntrain_epochs}
\end{table}
\newpage
\begin{table}[h]
    \centering
    \caption{Comparison of percentage $N_{\text{mem}}$ and $N_{\text{copy}}$ across different numbers of training samples when no. of training iterations are fixed.}
    \begin{tabular}{@{}lccccc@{}}
        \toprule
        \textbf{Dataset (\textbf{N$_{\text{itr}}$})} & \textbf{N$_{\text{train}}$} & \textbf{N$_{\text{mem}}$ (\%)} & & \textbf{N$_{\text{copy}}$ (\%)} \\
        \midrule
        \multirow{3}{*}{fastMRI (150k)} 
        & 1k & 28.4 & & 47.6 \\
        & 2k   & 8.6 & & 16.4 \\
        & 5k   & 4.2 & & 20.3 \\
        \midrule
        \multirow{3}{*}{Xray (100k)} 
        & 5k   & 54.7 & & 68.6 \\
        & 10k  & 11.2 & & 13.9 \\
        & 20k  & 5.7 & & 9.8 \\
        \bottomrule
    \end{tabular}
    \label{supptab:comparison_Ntrain_itr}
\end{table}
\newpage
\begin{table}[h]
    \centering
    \caption{Comparison of percentage $N_{\text{mem}}$ and $N_{\text{copy}}$ across different architecture sizes of diffusion models.}
    \begin{tabular}{@{}lccccc@{}}
        \toprule
        \textbf{Dataset (N$_{\text{train}}$)} & \textbf{Size} & \textbf{N$_{\text{mem}}$ (\%)} & & \textbf{N$_{\text{copy}}$ (\%)} \\
        \midrule
        \multirow{3}{*}{PCCTA (242)} 
        & Small  & 37.2 & & 79.8 \\
        & Medium & 38.8 & & 85.5 \\
        & Large  & 35.1 & & 76.0 \\
        \midrule
        \multirow{3}{*}{MRNET (904)} 
        & Small  & 21.9 & & 34.5 \\
        & Medium & 48.3 & & 87.4 \\
        & Large  & 45.8 & & 90.0 \\
        \midrule
        \multirow{3}{*}{fastMRI (1k)} 
        & Small  & 13.4 & & 15.9 \\
        & Medium & 30.6 & & 45.9 \\
        & Large  & 27.2 & & 44.8 \\
        \midrule
        \multirow{3}{*}{Xray (10k)} 
        & Small  & 5.1 & & 6.0 \\
        & Medium & 32.6 & & 54.5 \\
        & Large  & 30.9 & & 51.3 \\
        \bottomrule
    \end{tabular}
    \label{supptab:comparison_architecture_size2}
\end{table}
\end{document}